# Collective Motility, Mechanical Waves, and Durotaxis in Cell Clusters

Youyuan Deng[1,2], Herbert Levine[1,3], Xiaoming Mao[4], and Leonard M. Sander[4,5]

[1]Center for Theoretical Biological Physics, Rice University, Houston, Texas 77030-1402, USA
[2]Applied Physics Graduate Program, Rice University, Houston, Texas 77005-1827, USA
[3]Department of Physics, Northeastern University, Boston, Massachusetts 02115, USA
[4]Department of Physics, University of Michigan, Ann Arbor, Michigan 48109-1040, USA
[5]Center for the Study of Complex Systems, University of Michigan, Ann Arbor, Michigan 48109-1107, USA

January 20, 2021

## Abstract

Epithelial cell clusters often move collectively on a substrate. Mechanical signals play a major role in organizing this behavior. There are a number of experimental observations in these systems which await a comprehensive explanation. These include: the internal strains are tensile even for clusters that expand by proliferation; the tractions on the substrate are often confined to the edges of the cluster; there can exist density waves within the cluster; clusters can exhibit collective durotaxis when individual cells show no effect; and for cells in an annulus there is a transition between expanding clusters with proliferation and the case where cells fill the annulus and rotate around it. We formulate a mechanical model to examine these effects. We use a molecular clutch picture which allows "stalling" — inhibition of cell contraction by external forces. Stalled cells are passive from a physical point of view and the un-stalled cells are active. By attaching cells to the substrate and to each other, and taking into account contact inhibition of locomotion, we get a simple picture for many of these findings as well as predictions that could be tested.

## Author Summary

The collective motility of epithelial cell clusters plays a central role in biological processes during development, tissue repair and cancer invasion. We propose a computational model for the mechanics and dynamics of such clusters. This model starts from simple assumptions of how cell locomotion is affected by forces from neighboring cells and the substrate, and offers a unified picture to explain many of the most salient features observed in experiments from spatiotemporal patterns of stress in expanding cell clusters to mechanical waves in tissues and collective durotaxis. Our model unravels the role of mechanics in the motion of cell clusters, and provides insight into the collective motility of cells as an intriguing active matter system.

# 1 Introduction

Eukaryotic cells can often move by a judicious use of forces generated by their cytoskeleton and applied to their surroundings [1]. The observed motion can range from individual cells moving through extracellular space to the coordinated collective motion seen during developmental morphogenetic processes such as gastrulation. In fact, many processes that are important in biology and medicine involve the collective motility of epithelial cell sheets and clusters. In addition to morphogenesis, this type of motion is important during tissue repair, and cancer invasion [2]; for a recent review see [3]. A particularly striking example occurs as part of the progression of inflammatory breast cancer, where the rapid progress of the disease has been connected to collective cell motion [4, 5].

Aside from its direct biological relevance, the phenomenon of collective cell motion is of great interest from the perspective of non-equilibrium physics. Individual cells are active particles [6], able to use their stores of ATP to remain far from equilibrium, do work on their surroundings and on their neighbors, and more generally evade many of the features we associate with non-active materials. During collective motion, these cells coordinate their activity by mechanical coupling, for example by connections such as adherens junctions [7]. This coordination can further be modulated by signaling processes, helping to determine cellular front-back polarity [8] which affects the directionality of applied forces. How the interplay of all these effects give rise to the observed phenomenology is a challenging conceptual problem. In this paper we address the problem with a simplified mechanical model that helps explain many of the observed features. For simplicity we focus on one dimensional geometries (lines of cells moving on a substrate). Extension to two dimensions is conceptually straightforward albeit computationally expensive. Collective motion in 3d is more challenging as the cells do not have a convenient flat surface upon which to exert traction and will require physics beyond what is discussed here.

Collective motility has been studied in a wide variety of experiments, for a wide variety of cell types [9]. From the physics perspective, major progress has been made by utilizing convenient choices of cells, for example Madin-Darby Canine Kidney (MDCK) cells, moving on well-defined substrates that can be patterned by standard lithography techniques. Our primary concern relates to the physical forces between the cells, and these have been measured in several studies using traction force microscopy [10, 11, 12]. There is significant evidence from this body of work that the interaction between cells that produces collective behavior is primarily mechanical.

The mechanics of the clusters exhibits some odd features: for example in [10] it was shown that the mechanical stress in the center of a cluster is primarily tensile even though there is cell division and the cluster continually expands in size. In these experiments tension and cell density varied on the scale of millimeters. Conversely, in [12, 11] it is shown that the intercellular tension increased up to a plateau within a few cells of the boundary. In these newer experiments it was shown that most of the traction on the substrate comes from the outer parts of the cluster – in terms of net force applied, it is as if the center is barely attached to the substrate. This finding is most pronounced at early times, but persists to some extent even as the overall pattern begins to exhibit increasingly random fluctuations., Our model gives a plausible explanation for these behaviors, and shows that a key parameter is the rate of cell division. Specifically, we will see that increased division adds fluctuations to the cluster interior, disrupts the perfect confinement of substrate traction to boundary layers, and distorts the plateau shape of the interior tension.

There have been additional findings regarding collectively moving cells. Sometimes, mechanical waves are observed within the cluster [12, 13]. In the experiments of ref. [12], waves originate at the boundary following the release of the confluent layer from confinement. Traces of waves continue to exist until more than 10 hours after the release. In [13] there is spontaneous generation of repeated waves which the authors attribute to a linear instability of the system. We will propose below that these waves propagate due to the effective finite response rate of the cells once they are released from conditions that do not allow activity, but this release can be initiated via different mechanisms in differing experimental protocols. Most recently, an experiment which confines one-dimensional clusters of cells in annular rings [14] has shown a fascinating transition between growth with expansion and collective unidirectional motility without cell division. This phenomenon as well can be observed and explained within the framework we employ.

Individual cells can be sensitive to the stiffness of their environment, leading to the phenomenon of durotaxis [15, 16, 17, 18] where cells move up stiffness gradients. It is therefore interesting to consider whether cells moving collectively can exhibit increased stiffness sensitivity. Several papers have [11, 19] have investigated this type of collective durotaxis. In these cases individual cells exhibited negligible durotaxis, but the cluster did systematically expand more rapidly towards the stiffer side. Explanations that treat the cluster as as a giant single cell can explain some aspects of the data, but do not explain why the tissue is behaving in this manner. We have extended our model to treat durotaxis phenomenologically for the case of a stiff substrate and we indeed find collective durotaxis as long as the tissue continues its overall expansion. This expansion can either be due to transient effects after release or, over a longer time scale, due to on going cell division.

There have been a number of attempts to formulate theoretical models which can explain the forces in the cell clusters and concomitant motility, e.g. [20, 21, 11, 22, 14]. These models take a variety of forms. Some authors have modeled the cell cluster as a continuous active medium. In [20] the cluster is treated as a viscous fluid with an effective viscosity and friction coefficient which interacts with a nematic-like polarization field. Continuum models are also used to investigate questions regarding the stability of the advancing tissue boundary [23, 24, 25].

In [21] a continuum model is proposed for wave propagation using an assumed feedback between strain and an internal variable of the cell cluster. In [13] a continuum model for waves is given with a coupling between strain and polarity. The model is qualitatively compatible with the experiments for the case of instability waves. A treatment very close to the one we describe below was given by [14] for the transition to collective unidirectional motility in annulus. This treatment considers each cell to be in one of three different states, left polarized, right polarized, or stationary. Transitions between the states are governed by a master equation which takes into account contact inhibition of locomotion (CIL), an important mechanism for cells in close contact; see [3] and references therein. In the treatment described below we use the molecular clutch scheme to allow a more mechanistic model of the active motion of the cells in relation to their environment, and applied it to phenomena including waves and collective durotaxis.

At the other end of the spectrum are models which attempt to fully resolve the shape degrees of freedom of the individual cells. These include cellular Potts models [26], vertex models [27] and phase-field approaches [28, 29]. There has been only limited successes in using these models to study the detailed mechanical state of the cluster and the existence of the aforementioned waves. Finally, there are simplified cell approaches, ranging from the extreme of treating the cell as a single point [30] on upward to more complex collections of subcellular point-like elements [31]. In [22], cells are treated as composed of two force centers coupled by a contractile spring and which interact with other cells via adhesion forces. The theory includes cell proliferation and CIL. CIL describes the tendency for cells that collide to move away from each other. In the model outlined below we include these two effects. In [11] the cell monolayer as a whole is treated using a molecular clutch scheme [32] much like the one we propose in the work below for individual cells. As mentioned, however, the observed feature that the tractions are localized at the edges of the cluster "super-cell" was put in by hand in [11].

In the following we present a one-dimensional model for cells connected to a substrate by bonds that represent focal adhesions, and the internal dynamics is given by a version of the molecular clutch scheme given in [32]. Cells are joined by bonds and their motion is modulated by CIL. In our version of CIL, cells slow down to avoid hitting any barrier in front. Also, when two cells have a head-to-head collision, one or both of them (chosen at random) reverse polarization; in [22] a more general version was used. As will be explained in detail below, this polarization affects the distribution of adhesion sites, as is commonly seen in experiment [33]; adhesions are formed in the front and are disassembled in the rear. The dynamics of each cell undergoes a cycle of contraction and protrusion, as was originally proposed for single cells [34, 35]. What is new here is that the contraction is directly coupled to inter- and intercellular forces through a linear contraction speed-load relation modeling the clutch mechanism. Namely, if the cell tensile stress is too high, the cells will not be able to contract and will instead "stall". This is analogous to the stationary state in [14]. This notion is compatible with the observation in [11] and in [12] that the interior of the cluster sees small cell speeds/traction and large tension. As we will see below, this notion of stalled cells is key to explaining many of the observed features of cluster mechanics. These features, to be discussed in detail below, include the waves, the existence of collective durotaxis, and transitions from expanding clusters to collectively translating ones.

## 2 One dimensional model

In this section we describe a simplified, one-dimensional model for collective motility based on the molecular clutch concept [32]. This enables us to give a unified account of many mechanical features of epithelial cell clusters.

### 2.1 Cell Motility and the Molecular Clutch Model

The starting point for our model is the assumption of a motility cycle: cells contract and partially detach from the substrate by breaking adhesive bonds (more in the back than in the front), and the cell protrudes forward. Bonds can re-attach after cell protrusion. The cell then contracts again. We need explicit algorithms governing what happens to the cell position and to the forces during each stage of the cycle. In [22, 36], each cell is considered to be composed of two subcellular elements that interact with a pre-defined active contractile spring force law. Other work [34, 35] assumed that the cells have a fixed contraction speed during that part of the cycle. Both of these assumptions are rather simplified views of the complex process of myosin motor mini-filaments walking along actin fibers.

Here we use a variant of the molecular clutch model [32]. In this more realistic account, the molecular motors that drive contraction have a nontrivial force-velocity curve and thereby allow the cell to stall (i.e. pause contraction) when the tension applied to the cell is too large. As shown in Fig 1, we model the cell body as a contracting one-dimensional "bar" which is uniformly compressed around the mid-point by the contracting actin cytoskeleton. We take the forces generated by myosin motors to be concentrated at the midpoint, thus dividing the cell into front/back halves of equal lengths $L(t)$. The retarding force acting against contraction is generated by the adhesions to the substrate and the connections to the other cells; see below. This force is the same as the tension, $T$, at the cell midpoint, since the cell is in force balance. The condition for stalling is that $T$ is greater than $T_s$, the stall tension. In addition, we denote the force felt by each cell's front end as $F_h$ (h for head). The significance of this force will be relevant to our formulation of contact inhibition of locomotion (CIL), see next Section 2.2.

In a time step of length $dt$, the half-length contracts from $L$ to $L - dL$, where

$$dL = dt\, f(T) g(F_h) \tag{1}$$

The linear speed-load curve of molecular clutch gives

$$f(T) = \begin{cases} v_f & \text{if } T \leq 0 \\ v_f(1 - T/T_s) & \text{if } 0 < T < T_s \\ 0 & \text{if } T \geq T_s. \end{cases} \tag{2}$$

$g(F_h)$ is related to the CIL effect and hence for a freely moving cell, $g(F_h) = 1$. The full definition will be given in Section 2.2.

The cell starts each contraction cycle with half-length $L_0$. It then contracts for multiple steps according to Eq 1, before reaching the minimum half-length of $(1 - r_{contr})L_0$. Afterwards, it reverts to $L_0$ by protruding forward, and then enters the next contraction cycle. This picture of contraction/protrusion is completed by modeling the dynamics of cell's attachment to the substrate.

We assume a rigid substrate approximation. The cell is attached to substrate with elastic adhesion bonds, which describe trans-membrane proteins such as integrin. We represent these as a number of springs with rest length zero. Consider a cell whose midpoint has coordinate $x_c$. At the beginning of a contraction cycle, a series of springs is formed with one end on cell body at $x_i^{(c)}$ and the other on substrate at $x_i^{(s)} = x_i^{(c)}$, $i = -N_{adh,back}, -N_{adh,back} + 1, \ldots, -1, 1, 2, \ldots, N_{adh,front}$, and $N_{adh,back} < N_{adh,front}$, where negative $i$ indicates an adhesion in the back half, while positive the front. In this process, two adhesions of indices $i = -N_{adh,back}, N_{adh,front}$ form at both ends of the cell, and others are drawn by choosing the ratios from two groups of equally spaced probability distributions:

$$r_i^{(c)} = \frac{x_i^{(c)} - x_c}{x_{N_{adh,front}}^{(c)} - x_c}$$

$$\sim \begin{cases} \overline{\mathcal{N}}(\mu = \frac{i}{N_{adh,back}}, \sigma = \frac{1}{4N_{adh,back}}) & \text{if} \\ i = -N_{adh,back} + 1, \ldots, -1 \\ \overline{\mathcal{N}}(\mu = \frac{i}{N_{adh,front}}, \sigma = \frac{1}{4N_{adh,front}}) & \text{if} \\ i = 1, \ldots, N_{adh,front} - 1. \end{cases} \tag{3}$$

$\overline{\mathcal{N}}$ denotes the normal distribution truncated to within [-1, 1], so as to always lie within the cell body.

As the cell body contracts, the $x_i^{(c)}$ change, but the absolute coordinates $x_i^{(s)}$ are unchanged. Therefore, the $i$-th adhesion is stretched because $x_i^{(c)} \neq x_i^{(s)}$, and exerts a force $f_i = k(x_i^{(s)} - x_i^{(c)})$

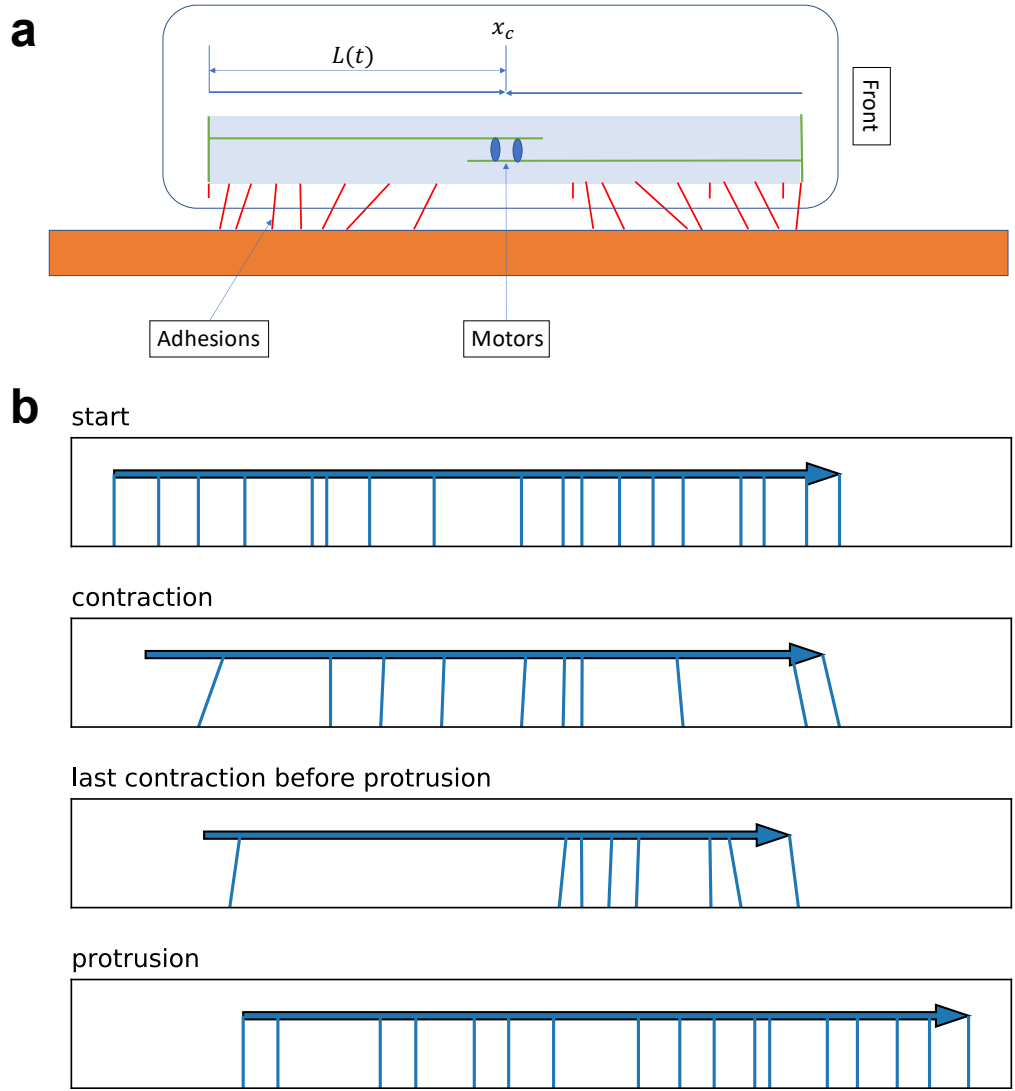

Figure 1: (a) One-dimensional model for a single cell. The red lines are springs with spring constant $k$, representing adhesions to substrate. The adhesions detach with rate $k_{off}$ and attach with rate $k_{on}$. The cell length contracts according to Eq 1. At the start of a contraction cycle, more adhesions form in the front half than in back. In the figure, the long bonds represent adhesions that attached to the substrate and the shorter ones represent ones that have detached. The heights of cellular components are for illustration only — the model is one-dimensional. (b) Snapshots of key steps during contraction cycles.

on the cell. The adhesions can detach when $f_i$ becomes large. We take the rate of detachment to be:

$$k_{off} = K \exp(f_i/F_d), \qquad (4)$$

where $F_d$ is the critical detachment force. The adhesions revert to zero length when detached. Afterwards, they randomly reattach with a constant rate $K$.

Mechanical relaxation is a much faster process than a typical biochemical reaction. In our model cells undergo immediate mechanical equilibration by shifting midpoint position $x_c$ after each biochemical change, i.e. each contraction and any detachment/attachment of adhesions. Since each half of the cell body is in equilibrium, the tension $T$ in Eq 1 must be equal to the total force exerted on the half-body by the adhesions with the substrate and with adjacent cells (to be discussed later):

$$T = \sum_{\text{half cell}} (f_i + f_{inter}) \qquad (5)$$

Returning to the discussion of protrusion, the cell is allowed to contract by a maximum ratio $r_{contr}$. Then it "protrudes" by reverting $L(t)$ to $L_0$ and placing the back end at the $x_i^{(c)}$ of the current rear-most adhesion, as in the models of [34, 35]. In a cluster of multiple cells, each cell is only allowed to protrude to occupy the inter-cellular space, and is prevented from overlapping with the neighboring cell, so they may protrude to a length smaller than $L_0$.

## 2.2 Contact Inhibition of Locomotion in Cell Clusters

With the speed-load function $f(T)$ in Eq 1, we capture the slowing-down of myosin-based contraction due to opposing forcers, but there is so far no inhibition of cellular motility when the cells face some barrier in front of them — it will contract as usual and protrude onto the barrier, despite the CIL effect. To resolve this artifact, we define a factor $g(F_h)$ following the form of $f(T)$ with the introduction of $F_{hs}$, the head-stopping force constant:

$$g(F_h) = \begin{cases} 1 & \text{if } F_h \geq 0 \\ (1 + F_h/F_{hs}) & \text{if } -F_{hs} < F_h < 0 \\ 0 & \text{if } T \leq -F_{hs}. \end{cases} \qquad (6)$$

where $F_h$ takes positive sign when aligned with cell back-to-front vector. If $F_h$ is solely exerted by cell-cell connection, positive sign corresponds to a tensile inter-cellular force.

To make a cluster in our model, the nearest ends of adjacent cells are joined by a spring with elastic constant $k$, and a fixed, non-zero rest length $l_0$. This elastic bond represents not only the separation between cells, but also the elasticity of the cell body. All adjacent cells are joined with springs of the same $k$ and $l_0$. A subtlety is necessary in expressing the force and potential energy of these inter-cellular springs. For an isotropic harmonic spring with rest length $l_0 \neq 0$, the potential energy is usually taken to be $V(\mathbf{x_1}, \mathbf{x_2}) \propto (|\mathbf{x_1} - \mathbf{x_2}| - l_0)^2$. That is, the spring is equally inclined to restore the natural length in either direction. To account for volume exclusion, the inter-cellular adhesion should not allow an equilibrium where two connected cells intrude into each other. A more sensible form for our model is $V(\mathbf{x_1}, \mathbf{x_2}) \propto |\mathbf{x_1} - \mathbf{x_2} - \mathbf{l_0}|^2$, with $\mathbf{l_0}$ being a vector.

Contact inhibition of locomotion (CIL) when applied specifically to cells in contact is a process during which cells actively alter their direction of movement to avoid collision, in addition to slowing down as in Eq 6. In our one-dimensional model, there are two possible polarities, left or right. The polarity is defined by the distribution of cell-substrate adhesions — the half with more adhesions is the front half; cells always protrude from the front half. When a spring connects front ends of two cells and is being compressed, a head-to-head collision is taking place. Contact inhibition would result in disassembly and assembly of adhesion complexes in front and back, respectively. Our model approximates this by randomly relocating colliding cells' detached cell-substrate adhesions in the front half to the back with rate:

$$k_r = K' \exp(f_{inter}/F_{CIL}), \tag{7}$$

excluding the $N_{adh,front}$-th one, i.e. one located at the front end. Suppose adhesion $j$ is chosen to move, we then randomly choose an interval in the back half delimited by two adjacent adhesions $j'$ and $j'+1$, where probability $P$(choosing j' and j'+1) $\propto l_{j'} = r^{(c)}_{j'+1} - r^{(c)}_{j'}$. The $j$-th adhesion then relocates to near the midpoint of $x^{(c)}_{j'}$ and $x^{(c)}_{j'+1}$: the new $r^{(c)}_j$ is drawn from $\overline{\mathcal{N}}(\mu = (r^{(c)}_{j'+1} + r^{(c)}_{j'})/2, \sigma = l_{j'}/8)$. See Eq 3 for the meaning of $r^{(c)}_j$ and $\overline{\mathcal{N}}$. Once the current rear half has more cell-substrate adhesions, the cell flips polarity, i.e. it protrudes from the end which now has more adhesions.

## 2.3 Cell Division and the Complete Algorithm

As in [22, 36], we adopt the idea that cells are likely to divide if the intra-cellular tension is large enough. At each step, if a cell's tension $T$ is greater than $T_{div}$, the critical tension, it divides with constant probability $dt\, r_{div}$. Upon division, a newborn cell C' of the same polarity is inserted next to the current cell C, randomly on the left or right. C' virtually protrudes in place to avoid overlapping (see the discussion in Section 2.1). The corresponding inter-cellular adhesion is cut to accommodate the new cell, and the nearest ends from adjacent cells are then reconnected. Note that the processes of applying CIL and of cell division are both significant configurational changes, and require mechanical equilibration immediately after each step.

The following is our complete algorithm for simulating collective motility:

- Start each cell with length $2L_0$. Initialize all the adhesions to be at their rest length.
- For each time step $dt$,

  1. For each cell, compute $T$ according to Eq 5; contract according to Eq 1; if the cell has reached $r_{contr}$, protrude; in rare cases when no adhesions remain attached to substrate, wait for next step. Adhesions are stretched. Equilibrate cell cluster by shifting $\{x_c\}$.
  2. For each cell, test for detachment of cell-substrate adhesions using Eq 4, i.e. detach with probability $k_{off}dt$. Move $\{x_c\}$ to maintain mechanical equilibrium.
  3. For each cell, attach the free adhesions with probability $Kdt$. Equilibrate.
  4. For each cell, apply contact inhibition of locomotion (CIL). Equilibrate.
  5. For each cell, test for cell division, i.e. divide with probability $dt\, r_{div}$ if $T \geq T_{div}$. Equilibrate.

Fig 1(b) illustrates several key steps during such a contraction cycle. Model parameters are listed in Table 1.

| Symbol | Meaning | Value |
|---|---|---|
| $L_0$ | cell's (maximum) half-length at the beginning of each contraction cycle | 5 $\mu m$ |
| $v_f$ | cell's free(maximum) contraction speed, w.r.t half-length | 5 $\mu m$/min |
| $r_{contr}$ | cell's maximum allowed contraction ratio | 20 % |
| $T_s$ | cell's stall tension | 10 nN |
| $F_{hs}$ | cell's head-stopping force | 1.5 nN |
| $l_0$ | rest length of inter-cellular adhesions, also the initial inter-cellular separation except for the pre-confinement modelling | 5 $\mu m$ |
| $k$ | spring constant of cell-cell and cell-substrate adhesions | 1 nN/ $\mu m$ |
| $K$ | reattachment rate and coefficient in detachment rate expression of cell-substrate adhesion | 10 /min |
| $F_d$ | critical force for detachment of cell-substrate adhesions | 0.75 nN |
| $T_{div}$ | threshold tension of cell division | 0.99 $T_s$ |
| $k_{div}$ | rate of cell division once $T \geq T_{div}$ | 1 /min |
| $N_{adh,back}$ | number of adhesions to substrate in back half | 8 |
| $N_{adh,front}$ | number of adhesions to substrate in front half | 10 |
| $dt$ | time step size | 0.01 min |

Table 1: Parameters in one-dimensional model

# 3 Simulation Results

## 3.1 Cluster Dynamics in the Absence of Cell Division

Several previous models [34, 35] describe a crawling cell's motility cycle, but do not use the molecular clutch picture. In a reduced "cluster" consisting of a single isolated cell, the previous models are the limit of $T_s \to \infty$ of our current approach. (For an animation of a single free-moving cell, see SI Movie S1.)

To start to look at collective effects, we simulated a cluster consisting of two cells, aligned head-to-head. As the simulation starts, the two cells begin to collide. Because of the CIL mechanism, at least one of the two cells will eventually change its polarity. When one cell flips, the two cells will move together as a translating cluster. Due to the finite time step size, both cells may flip at the same step, leading to a static situation. See SI Section 1 and SI Movies S2, S3 for more details. We will see that these two basic choices, a static cluster with an equilibrated tug-of-war versus a translating state, also characterize multicellular clusters.

In order to represent a large cell cluster, we repeated the procedure above for 50 cells connected by springs with random initial polarity, and used the dynamics described in Section 2. In Fig 2, we show results from one simulation for the polarity, cell tension, force between cell and substrate, and inter-cellular tension. Note that after initial transients the colony settles down with large domains of like polarity (Fig 2(a), SI Fig 6(a)) pulling on each other. This is one of two possible outcomes; the other is that the large majority of cells move in one direction and the whole cluster translates.

For our initial conditions with random polarity for 50 cells, both outcomes occur with roughly equal likelihood. In SI Section 2 we show that the tug-of-war state is increasingly likely as the number of cells increases. The presence of an initial confinement period, as in most of the experimental protocols, also probably biases the outcome in favor of expansion rather than translation. We will discuss that initial condition later. Here we consider simulations that lead to tissue expansion with the majority of the cells on the left moving left and those on the right to the right. In the case shown in the figure, we have domains of similar size.

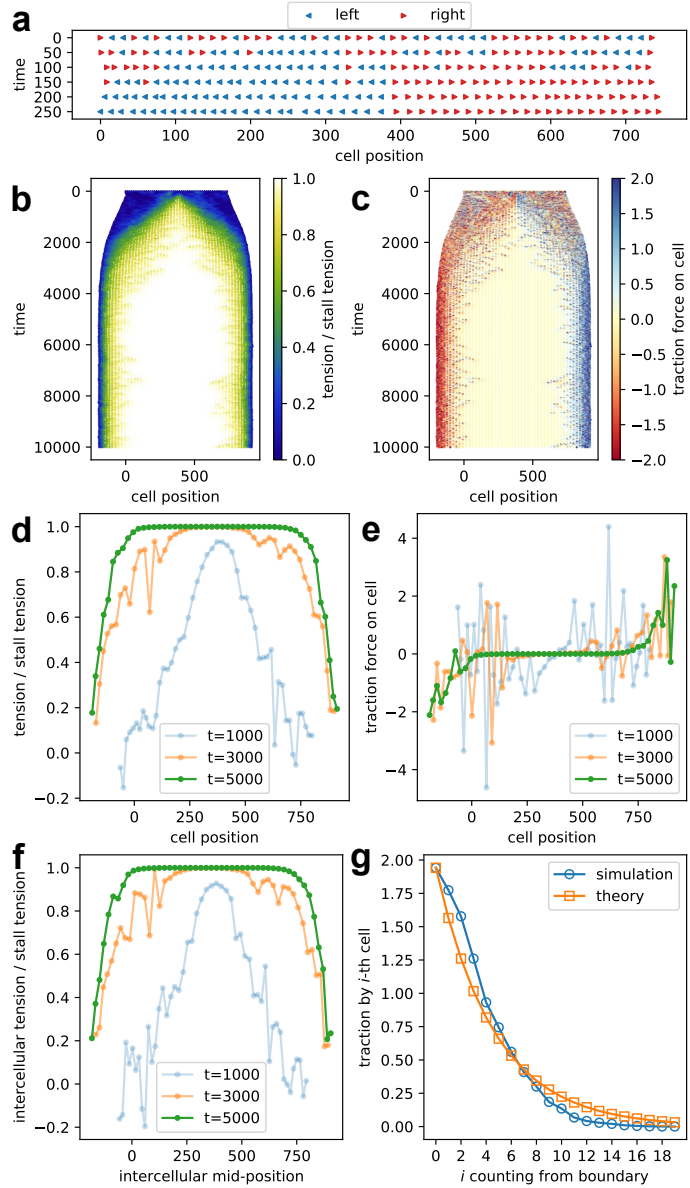

Figure 2: A simulated cell cluster without proliferation. (a) Polarity of the cells at select time steps during initial transient. They eventually form two similar-sized domains. (b) Kymograph of tension at each cell's midpoint. Note the stalling of the interior. (c) Kymograph of traction force on each cell by substrate. The kymographs shown here are composed of discrete points in space-time coordinate systems where each point represents a cell or an inter-cellular spring at a specific time-step. (d)(e) Same quantities as (b)(c), at select time steps. Note how the interior cells gradually become stalled. (f) Inter-cellular force at select time-steps, the main source of stalling tension in the interior. (g) Average traction on the left-most cells during the latter half of the trajectory, and predicted values from the simple theory discussed in the text.

There are interesting features in Fig 2 that closely correspond to experimental observations. Once the expansion slows, the traction force becomes confined to the edges of the colony, even though all of the cells are attached to the substrate (Fig 2(c, e)). This is because the interior cells are mostly stalled (not contracting) and the forces on either end of each domain balance (Fig 2(b, d)). Only at the edges are the cells pulling outwards. These traction forces at the edges eventually transmit stress to the interior via cell-cell junctions, which are therefore under large tension, so large as approaching $T_s$, the stall tension (Fig 2(f), SI Fig 6(d)). Namely, the interior cells are attached to substrate, but are not generating traction as they are stalled by large inter-cellular tension (as can be seen in Eqs 1, 2, and 5), which in turn originates from non-stalled active edge cells. Also note that the inter-cellular springs are tensile except for early transients [12, 11].

Traction forces vary throughout contraction cycles, and fluctuate due to the stochasticity of cell-substrate adhesions, so we averaged the traction by each cell over the later half of the trajectory; see Fig 2(g). One can read off from Fig 2(d–g) that intra- and inter-cellular tension accumulates from the outermost cells inward, while the traction exerted by each cell monotonically approaches zero. We have shown that the accumulation leads to stalling of the interior, but why is such monotonic behavior seen within the active edge layers? It is sensible to assume that each cell's net traction is made possible by protrusion, without which the forces from different cell-substrate bonds of the same cell cancel each other. Since the protrusion algorithm prevents overlapping, one might be tempted attribute this to the fact that inner cells tend to protrude less than outer cells. This however gives the wrong result because inter-cellular space increases inward with inter-cellular force. A plausible explanation lies in the molecular clutch speed-load curve. Roughly speaking, more frequent protrusion, i.e. shorter contraction cycle, leads to larger average traction. Thus, the mean value $\langle F_{trac} \rangle \propto 1/T_{cycle} \propto \langle \text{contraction speed} \rangle \propto \langle f(T) \rangle$. We index the cells with $i$ starting from 0 on the outside. Then we further approximate $T(i\text{-th cell}) = \sum_{i'<i} F_{trac}(i'\text{-th cell})$, and calibrate the proportionality coefficient using the 0-th cell (i.e. setting $\langle F_{trac} \rangle$ to the simulation value), we obtain the "theory" curve in Fig 2(g), which is comparable to the observed simulation values.

Given the curve in Fig 2(g), it is clear that a minimum number of active edge cells are needed to accumulate stress to reach stalling. On the other hand, stalled, non-contracting cells are effectively passive. Their adhesions to the substrate still randomly detach/reattach, amounting to an effective viscous friction. The friction is similar for active cells, but it is the only cell-substrate interaction for stalled cells. Of course, the viscous friction force is zero unless the cells actually moves. It is therefore natural to speculate that the polarity of stalled cells are irrelevant, so we can have force-balanced non-translating clusters, with unequal-sized polarity domains, so long as each domain contains more than the required minimum number of edge cells. Fig 3(a–c), SI Fig 7) shown an example of this type of behavior. On the other hand, when the left or right domain has too few cells, the accumulated tension is not enough and the interior cells are not stalled, but they still have uniformly weaker contraction compared to edge cells (Fig 3(d–f), SI Fig 8). In this case, the whole colony translates in bulk. The limiting case of course is when all cells end up having the same polarity (see SI Figs 9, 10). As discussed above, starting from a completely randomized polarity state allows both types of solutions to emerge dynamically.

The localization of traction and motility shown in Figs 2, 3 qualitatively agrees with experimental findings [12, 11]. This agreement relies on the fact that cells in the center are effectively stalled by tensile stress. Note however, that as a function of time in our simulation, the width of the cell colony saturates due to the equilibrium between the traction forces at edges and the interior forces on individual cells; on the other hand, those experiments observed continued expansion for ∼10 hours. This is, in our view, connected to the increasing importance of cell division, which occurs more frequently when cells are subject to mechanical stretching [37]. We will consider this effect below. The "head-stopping" aspect of CIL can lead to similar localization for confined clusters, as we will see below.

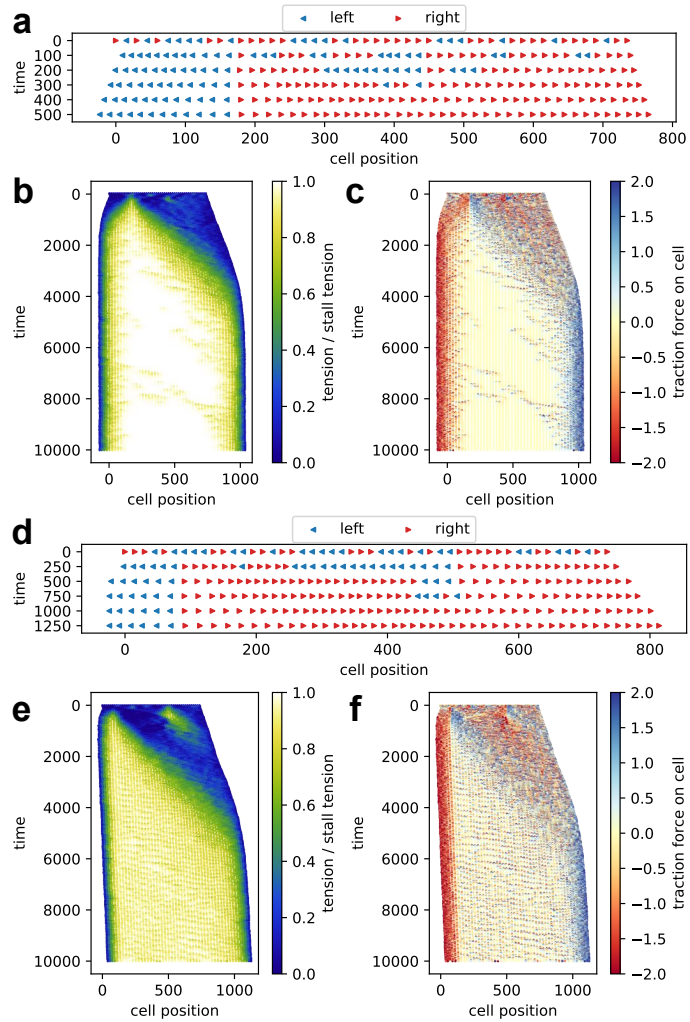

Figure 3: (a–c) Polarity at selected times and kymographs of tension and traction for a static cluster consisting of two unequal-sized domains. (d–f) Same quantities for a moving cluster where the left domain has too few cells to stall the interior.

## 3.2 Pre-confinement, Mechanical Waves and Cell Division

A common experimental procedure for studying tissue expansion is that cells are first confined within a rectangular stencil before that barrier is removed and cells are allowed to expand into a free zone [11, 12]. Subsequently, cells became mobilized progressively inward, and consequently mechanical quantities such as tension exhibit a wave-like pattern on the kymographs [12]. To our knowledge, there has been so far no consensus on the exact nature of these waves. In our interpretation, the early-time waves that initiate from both boundaries, travel inward, and cross each other, are related to the sudden release of the confining barrier. In experiments there are symmetrically and asymmetrically expanding clusters [12], which have already been captured by the randomness in initial polarity leading to different domain sizes. Let us we focus on an initial condition of 50 cells in two equal domains, with 25 cells on left and right with left/right polarity, respectively (Fig 4(a)).

It is not surprising that there is a finite time delay before the influence of barrier removal reaches the inner cells. In general, one should expect a finite relay speed of mechanical response in cell colonies. Such a delay would appear as a down-pointing triangle on a kymograph. We hypothesize that the prolonged initial confined growth induces compression of the cells, the release of which then leads to the crossed waves. To show this, we placed harmonic potentials acting as walls on both sides and changed the rest length of inter-cellular springs such that all the cells are under compression that is large enough so cells basically are stopped (recall Eq 6). At the 2000-th step, the two walls are removed. As shown in Fig 4(b), because of the mechanism in Eq 6, cell mobilization progresses gradually inward, and the tension begins to accumulate from center outward. In Fig 4(c) it is shown that the blue and red colors exchange position once they meet at the center at around the 3000-th step, so there is crossing rather than bouncing-back or reflection. From this perspective, it is surprising that the in the experiment the two arms of the triangle reflect off the edge and continue back to the opposite side [12]. Our model does not show these echos — they dissipate after reaching the opposite boundary. This may indicate a longer time influence of the original confinement than is present in our treatment.

We now turn to the effects of cell proliferation, using the scheme outlined in Section 2.3. Due to our CIL rules, there is no polarity reversal throughout the simulations in this section (Fig 4(a)). Now the clusters grow indefinitely as long as the critical division tension $T_{div}$ is smaller than $T_s$. Interestingly, there is now a new source of wave-like excitation, now launched from cell division sites. (Fig 4(b)(c)). Specifically, whenever a new cell is added, there is a strong local density perturbation and this appears to launch a density wave in the cluster which then propagates to the boundary. Note that this effect arises naturally in our model: we do not need an extra feedback mechanism as suggested in [21]. We understand the waves as arising simply because there is a time delay for a cell to start to move from its stalled state to accommodate the presence of the new cell. These waves are better separated when division is infrequent (Fig 4(d)), and are more overlapped when division is more frequent (Fig 4(e)). In Fig 4(d), it is clear that each wave arm consists of an upper edge where cells sequentially un-stall, and a lower edge where cells restore stalling. We should note however that it is unknown whether one can see these waves in one-dimensional projections of data from 2d experiments. Even though there is a bias for cell division and hence wave launch to occur near the center (see below), the initiation times for one dimensional slices are unlikely to be synchronized. This issue awaits elucidation by extensions of our current model to the two-dimensional case.

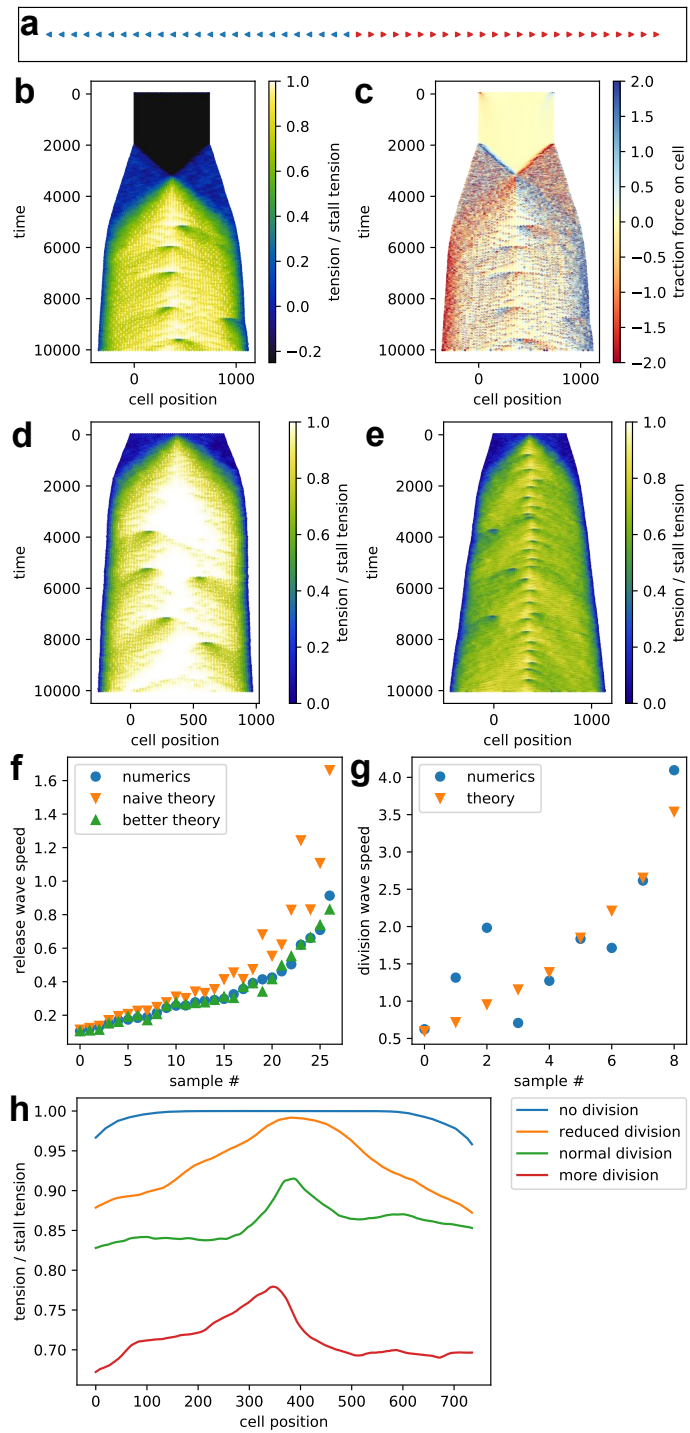

Figure 4: Waves in cell clusters due to confinement release and due to proliferation. (Caption continued on a different page.)

Figure 4: Waves in cell clusters due to confinement release and due to proliferation. (a) Initial setup of the cluster. (b)(c) Cell midpoint tension, traction kymographs of a proliferating cluster with pre-confinement, where $l_0$ is increased to 10 from that in Table 1. See also SI Fig 11. (d)(e) Cell midpoint tension kymographs for a (d) less / (e) more frequently proliferating cluster. See also SI Figs 12, 13, 14. (f) Release wave speed measurements and estimations. 27 samples were generated with varied $v_f$, $F_{hs}$, $l_0$. (g) Division wave speed measurements and estimations. 9 samples were generated with varied $v_f$ and initial separation between adjacent cell midpoints $\Delta x_c^{(init)}$. For the exact parameter variations in (f)(g) see SI Section 5. (h) Cell midpoint tension averaged over the latter half of a trajectory for different division frequency. The exact division-related parameter variations can be found under SI Fig 4.

In addition to generating the waves, proliferation events change the tension distribution across the cluster. As shown in Fig 4(h), the tension has a clear plateau shape in the absence of division. As division becomes more frequent, the average interior tension decreases and the peak at the center is sharper. Due to the peaked distribution, cell division events are more likely near the center, where intra-cellular tension reaches $T_{div}$ more readily. Although such a division would cause stress relief, the rapid propagation away from the initiation point of the waves quickly restores the center region to being the most tensile. This behavior could account for the observation of [10] where the tension gradient is not confined to the surface layers; we note that the experimental data comes from a 2D system and represents an average over some distance in the longitudinal direction, and this may smooth out the structure as compared to our 1D simulation results. These division events are also a source of noise in the interior. Since the interior is now not completely stalled, there is fluctuating, nonzero traction with the substrate from inner cells (Fig 4(c)), as compared to Fig 2. As we will see in Section 3.4, this noise is crucial for the system's dynamic response to stiffness gradients in the substrate.

It is worthwhile to compare these two different wave phenomena seen in our model. We relate the early-time wave pattern to the response to pre-confinement, and late-time to accommodation to cell divisions. Both are attributable to the inherent finite response to perturbations which alter the cell's motility from a state in which that motility was suppressed. The former type of wave initiates from the boundaries and propagates inward, while the latter initiates at the center and propagates outward. To demonstrate the underlying physics, we can look at the propagation speeds for both wave types. For the release wave, the wave speed measures how fast successive cells are "activated" one by one. Label the cells from outside as 0-th, 1-st, and so on, and consider the time needed between the sequential activation of $i$-th and $(i+1)$-th cell. The distance the wave travels is the initial separation between adjacent cell midpoints $\Delta x_c^{(init)}$. We approximate the $i$-th cell as traveling at the speed of single free cell speed $v_1$, which can be easily calculated from a single cell simulation. It needs to travel for a distance of $l_0 - (\Delta x_c^{(init)} - 2L_0) - F_{hs}/k$ (See Table 1 for parameter definitions) before the compression on the next cell's front is less than $F_{hs}$ allowing the next one to be activated. This gives a "naive theory" of wave speed

$$\frac{\Delta x_c^{(init)} v_1}{l_0 - (\Delta x_c^{(init)} - 2L_0) - F_{hs}/k}$$

(see Fig 4(f)). An improvement is made by considering that each cell except for the 0-th one linearly accelerates from zero to $v_1$. When the $(i-1)$-th cell's displacement is between $l_0 - (\Delta x_c^{(init)} - 2L_0) - F_{hs}/k$ and $l_0 - (\Delta x_c^{(init)} - 2L_0)$, the $(i-1)$-th travels at $v_1$, but the $i$-th cell travels at mean speed of $v_1/2$ because of the linear acceleration. That is, the $i$-th cell travels at mean speed $v_1/2$ for a duration of $F_{hs}/(kv_1)$, then travels at $v_1$ for the rest distance of $l_0 - (\Delta x_c^{(init)} - 2L_0) - F_{hs}/k - (v_1/2)(F_{hs}/(kv_1))$, before calling up the $(i+1)$-th cell. This gives a "better theory" of wave speed

$$\frac{\Delta x_c^{(init)} v_1}{l_0 - (\Delta x_c^{(init)} - 2L_0) - F_{hs}/(2k)}$$

(see Fig 4(f)). For the division-launched wave, note that the influence of insertion of the new-born cell is transmitted most strongly when the nearest cell to the division site protrudes, so one can estimate the wave upper edge (sequential un-stalling) speed to be $(l_0 + T_s/k + 2L_0)/T_{cycle}$, namely the distance between midpoints of adjacent stalled cells divided by time length of a contraction cycle. Note that $T_{cycle}$ is inversely proportional to $v_1$ and can be similarly calculated from a single-cell simulation. Given the diffuse nature of these division waves, the measurement can be hardly accurate, but this theory still approximately agrees with numerics (Fig 4(g)). In principle, one can use these relations to distinguish different wave types in real experiments.

The waves observed in [13] seem to be different from what we have discussed so far. The authors interpret their wave observations as arising from a spontaneous instability in their system which gives rise to repeated wave launches, presumably arising from the amplification of fluctuations. We have not observed such an instability in our simulations. We believe that the instability of [13] arises from a process that we do not have in our model. We can see this by examining their continuum theory. The process that gives unstable behavior is that the mean propulsive force of cells increases with strain. The underlying process seems to be that in a two-dimensional layer, uniaxial strain will align cells. Then a velocity fluctuation will cause additional strain which aligns more cells, giving positive feedback. Of course, our one-dimensional simulation cannot support such a process. In our model cell contraction and protrusion play the role of propulsive force, and in the molecular clutch scheme, Eq 1, contraction slows down as strain increases. For a real system, it is plausible that both effects might occur. Which one dominates probably depends on parameters and cell density. We should note that in [10, 12] there is no sign of an instability.

## 3.3 Periodic Boundary Conditions

In the experiment of [14], cells move along a 1D annulus. Initially, clusters expand but once the ends contact each other around the annulus, there is a transition between a state with expansion with proliferation and one with collective motility (rotation) without cell division. To treat this case, we simulated a cluster growing in a 1D periodic domain, ignoring any possible effect due to ring curvature. An extra intercellular spring between the two outermost cells in our colony is added when the cluster has expanded enough to "fill the annulus". Specifically for this simulation, this size occurs at the 5000-th step and thereafter the left- and right-most cells are joined by an adhesion; see Fig 5. Note that in these simulations we have enhanced the rate of cell division by taking $F_{div} = 0.9\ T_s$ (instead of $0.99\ T_s$) to speed up cluster growth. As can be seen, our simulation directly captures the observed transition.

The mechanism underlying the transition is that when the two outer ends of the cluster collide as the cells fill the annulus (i.e., when the new spring is attached), the CIL process becomes active. To capture the transition details, we plotted the polarity and cell length in a kymograph (Fig 5(a)). In our simulations the cluster always chooses one or the other polarity, and starts to revolve around the annulus. The resulting colony remain weakly tensile (Fig 5(b)(d)). Recalling that the condition for CIL re-polarization requires collision; individual cell(s) may not immediately orient so as to agree with the majority polarity, but are nonetheless dragged along (See the red segment after the 5000-th step in Fig 5(a)). The reversal of polarity takes place in a wave (the sloping border between red and blue in Fig 5(a) and the corresponding "scar" in (b–d)). The nature of this wave is similar to that of the density waves discussed above, involving finite delay in response to mechanical perturbations. There is a characteristic time for reversal of polarity, the inverse of the rate in Eq 7. The speed of the reversal wave is of the order of the cell separation divided by this time. Finally, the transition to rotation may not be so smooth. In some cases, the domain border is zigzag shaped on the kymograph. See SI Figs 15, 16, 17 for examples.

## 3.4 Collective Durotaxis

Durotaxis is motion directed by stiffness gradients [15, 16, 17, 18], where cells that are sensitive to such gradients move towards stiffer regions. The general understanding of this phenomenon in the literature relies on one of two possible mechanisms. One approach assumes that cells move up gradients because they move faster on stiff substrates, so that if they wander in to such a region they will wander away [16, 17]. This is sometimes called population durotaxis because it does not operate at the level of a single cell. A different mechanism assumes that cells can locally sense stiffness [38] and tend to grow stronger adhesions on stiff substrates. This leads to durotaxis at the single cell level [35] because the end of the cell in the stiff region (usually the front) will not detach as quickly as the end in the compliant part, leading the back to peel off and leading to motion up the gradient. Doering et al. [18] have argued that the latter mechanism is more likely for single cells.

Interestingly, in some cases, single cell durotaxis is not observed, but a large cluster does move up the stiffness gradient [11, 19]. Note that in the experiments, the cluster does not "translate" as a whole, but rather has different rates of expansion on two edges and hence its center of mass undergoes durotaxis. The interpretation for the effect given in these references is that the cells in the interior of the cluster are completely disconnected to the substrate, so that the cluster acts as a giant cell which spans wider zones with larger stiffness difference than what any single cell can sense. We will refer to this as the "giant cell" model'. Note however that in these experiments, symmetric expansion is sustained for ∼10 hours, and hence proliferation should contribute significantly to the expansion. Yet, we have shown that proliferation is a major source of interior noises, leading to nonzero and fluctuating interior traction, making the giant cell model unrealistic. As we will now discuss, our model predicts that non-trivial traction in the interior is integral to durotaxis.

Within our model, there are two mechanisms that we can invoke. We note that a spring connected to a more compliant substrate shares load and becomes, in effect, a softer spring. This suggests making the spring constant, $k$, space dependent, with larger $k$ for stiffer substrates. Note that larger $k$ should lead to faster cell speed since it makes detachment faster (recall Eq 1 and see SI Section 4). This mechanism would give population durotaxis if all the cells were mechanically uncoupled, corresponding to the first mechanism above. Alternatively, we can try to account for the effects of forming more and stronger focal adhesions on stiffer surfaces by letting $F_d$ depend on space such that $F_d$ is larger in stiffer regions. A larger $F_d$ makes it harder to detach thus making the adhesions more numerous. Based on the experimental results, the possibility of single-cell level durotaxis is not considered in either approach; specifically, the now varying $k$ or $F_d$ is taken to depend on cell midpoint coordinate, $x_c$, and is uniform within individual cells. The major effect left out of our treatment is the relaxation of the substrate under the cluster. A model resolving substrate degrees of freedom is included in [35], but in this context the necessary computation is also at least two orders of magnitude more time-consuming. For a relatively stiff substrate it is reasonable to assume that this substrate relaxation effect is not qualitatively important. That is to say we take the limit where the value of the substrate stiffness goes to infinity, but the difference in stiffness between the two ends of the cluster is fixed. Note that in [39] substrate relaxation is also neglected and in [40] a purely local spring softening is assumed without solving globally for substrate deformation.

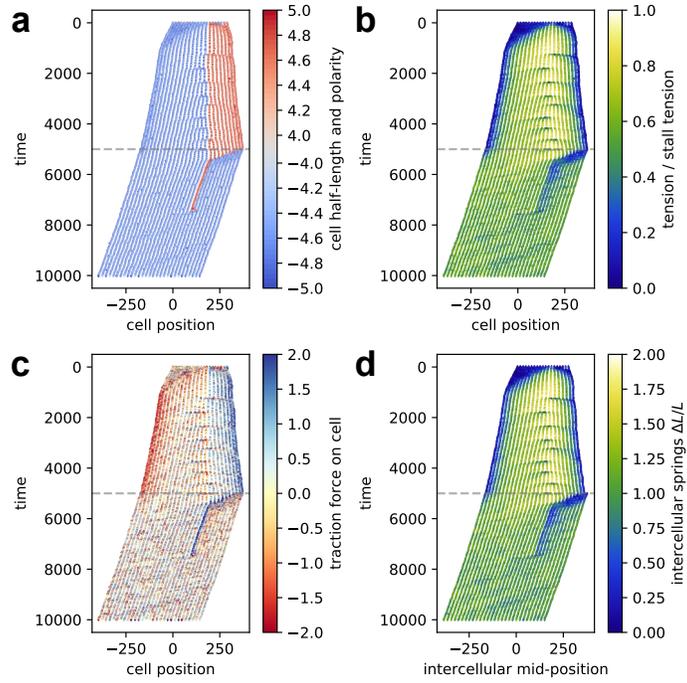

Figure 5: Space-time plot of a simulated cell cluster on a ring. It keeps expanding until two outermost cells are joined by a spring at t = 5000, indicated by the gray dashed line. (a) Polarity and half-length of the cells. Blue or negative sign denotes "left" polarity and red/positive denotes "right". (b) Cell midpoint tension. (c) Traction force on each cell by substrate due to the adhesions. (d) Inter-cellular tension stress, i.e. the stretch of the inter-cellular springs. Note that the rightmost new spring created at $t = 5000$ is the one connecting two outermost cells.

In Fig 6 and SI Fig 18 we show the results. As in Section 3.2, we started with 50 cells with two equal-sized domains. An average shift in the center of mass is calculated from 100 independent samples for each mechanism/parameter. There is sustained collective durotaxis for either mechanism when proliferation is on (Fig 6(a)(c)). When proliferation is turned off, the clusters also shift in the right direction but only during initial transient expansion (Fig 6(b)(d)). That durotaxis occurs only during expansion (transient or sustained) is consistent with the experimental picture mentioned above. As expected, as the gradient of stiffness increases, the the center of mass translation of the clusters increases, though a tendency for saturation seems to appear.

As one can see from Fig 6(b')(d'), when a steady state is established in the absence of proliferation, there is again a saturation of cluster size and equilibrium between traction forces at both edges; the interior is again mostly stalled; cf. Fig 2. As has been argued in Section 3.1, stalled cells are purely passive agents having only frictional interaction with substrate. How this friction depends on stiffness is a different question, but the non-proliferating clusters always remain static without external perturbations, so the actual viscous friction is always zero. What is notably different from Fig 2 is that the left active edges of Fig 6(b')(d') are thicker than the right, meaning the traction activity is more diffuse near the left edges where the substrate is more compliant. Given the equilibrium of traction forces, this indicates that each of the left edge cells on average generates weaker traction than the right ones, so more cells stay active to compensate. Overall, the non-proliferating clusters can sense the stiffness variation, but a dynamical response in the form of durotaxis is only possible during initial transient. Once the cluster reaches the saturation width, no overall translation is possible. In the cases with proliferation, a similar difference in the diffuseness of edge traction can be seen in Fig 6(a')(c') despite the noise introduced by the division events. Similar to what was seen in the uniform substrate cases in Section 3.2, Fig 4, the interior traction in Fig 6(a')(c') is nonzero as a result of proliferation, although more traction is still generated near the edges. Overall, the proliferating clusters can continuously durotax only because cell division events "shake up" the system for it to respond to stiffness gradient dynamically. Recall that the experiments cited above were for proliferating monolayers. These cases are consistent with our model. It would be interesting to look for collective durotaxis in epithelial clusters where proliferation is suppressed. Finally, note that all the above features are similar between Fig 6(a, a', b, b') and (c, c', d, d'), i.e. between the two underlying mechanisms, bolstering the robustness of our results.

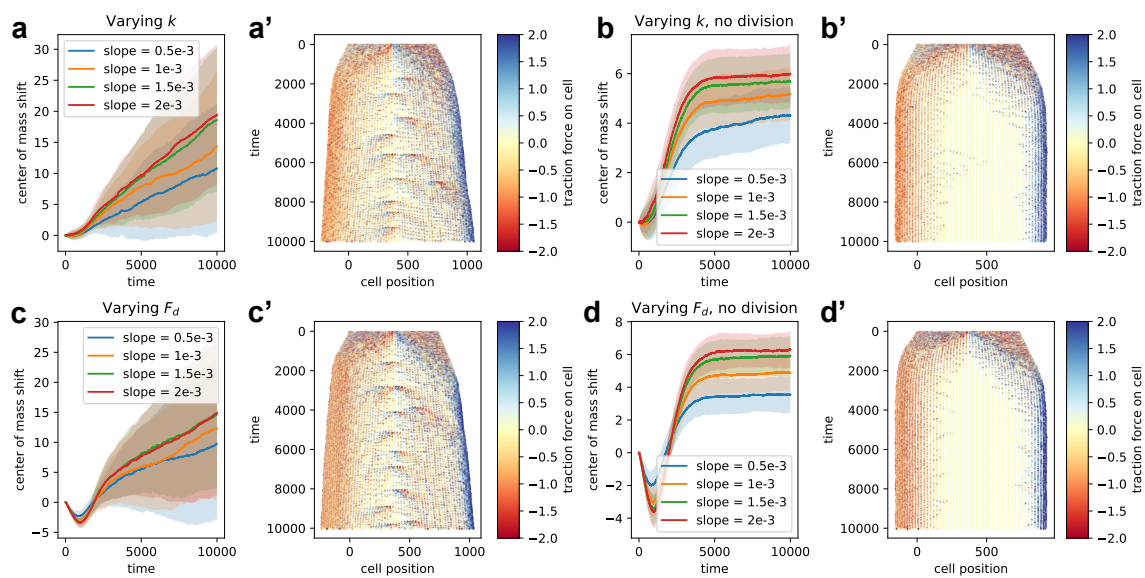

Figure 6: Durotaxis: The necessity of cell division. (a–d) Center of mass shift of the clusters as a function of time, for two effective stiffness mechanisms and different gradients of stiffness. Each line is the average over 100 samples, and the filled area signifies the standard deviation. (a)(b) Space-dependent elastic constant $k$: $k = \max(0.2, 0.4 + \text{slope} * x_c)$, with proliferation on (a) / off (b). (c)(d) Space-dependent $F_d$: $F_d = \max(0.2, 0.4 + \text{slope} * x_c)$, with proliferation on (c) / off (d). (a'–d') Cell traction kymographs of a sample with slope $1.5e-3$ from each of the groups (a–d).

# 4  Summary and Conclusions

Here we have introduced a simple mechanical model for cells that are attached to each other by molecular springs and hence move collectively. This model is based on the notion that cells undergo a contraction-protrusion cycle and that these processes are modulated by adhesion to the substrate and by intercellular forces. To account for the fact that contraction is based on myosin mini-filaments walking along actin fibers, we have used the molecular clutch formulation of the connection between the stress state of the cell and the contractile velocity. Adhesion is treated as a set of springs connecting points along the cell to a rigid substrate. These adhesive springs come and go and cell polarity determines which half of the cell has a higher number of such adhesions. We have also incorporated a simplified form of the well-established biological mechanism CIL, contact inhibition of locomotion. This form of CIL consists of slowing down the cell when it encounters an obstacle and also reversing the polarity by moving adhesive sites of two cells engaged in a head-on collision. This model is formulated in one dimension, i.e. for a moving line of cells.

Given its relative simplicity, it is remarkable how many interesting aspects of collective cell motility this model is able to encompass. As shown in detail, the eventual mechanical state of a cell cluster can be of one of several types. In the absence of any cell division, the cluster size eventually must saturate. If the cluster is relatively symmetric, that is there are a significant number of cells in the right-polarized domain engaging in a tug-of-war of with a significant number of left-polarized cells, the cluster will stop moving altogether and the intercellular tension will exhibit a broad plateau. This is similar to the mechanical state observed in [12, 11], though in those experiments there is proliferation. The plateau region is composed of a large number of cells that have stalled and hence are no longer actively contracting. The tension at the plateau then falls to zero over a finite size region at the two edges. Interestingly, there is no translation of the cluster even if the number of differently polarized cells are unequal; this is because the number of actively pulling cells is the same on both sides and the different numbers of stalled cells make no difference. A different possibility is that there is polarity pattern width with only a small number of cells in the minimum polarity direction, smaller than the transition region width. The limiting case here is when all the cells are polarized in the same direction. Then the cells are never stalled, and the entire cluster moves systematically.

In order to allow the cluster to grow in size without saturation, we include cell division. It has been argued that cell division is directly coupled to the size of cells [31] which in our model is directly determined by the tensile stress. we have therefore allowed any cell to divide if its tension gets close to the stall value. For small rates of cell division, the previous "plateau" state is relatively unchanged except for the fact that it continues to slowly expand, simply by adding more stalled cells to the cluster interior.

Our model can explain the occurrence of waves emerging from the cluster boundary once it is released from confinement and allowed to expand. These waves have been observed in several experiments on systems of this type and have been attributed to a variety of complex mechanisms [41, 21]. In our model these results come simply from the time it takes fro a cell to recover motility from conditions under which it is suppressed. A new aspect of this phenomenon is observed here; cell division events also each lead to a propagating disturbance, moving faster than the expansion rate and hence hitting the cluster boundary and dissipating. Again, the propagating disturbance is simply due to the transient release from stall due to the local compression created by the division event. As argued above that in a two-dimensional system these disturbances may be difficult to observe, but they do have observable effects: they can give rise to traction disorder within the cluster as in [12]. As the waves relax the stall condition, cells in the interior undergo active contraction and hence contribute to the net traction force. This tends to destroy the plateau and spreads the tension gradient region over the entire cluster. This type of pattern is perhaps similar to what was observed in [10], where the tension gradient exists over several millimeters worth of tissue. The wave effect offers an alternate explanation for this behavior than that provided in [22] which assumed that cells never reach stall forces anywhere inside the tissue.

The rearrangement of the cells caused either by transient expansion or in steady-state due to proliferation is necessary for the collective durotaxis seen in our extended model. This prediction could be directly tested experimentally. We treated this problem approximately in that we do not directly consider the substrate deformation. This is presumably reasonable in the case of a very stiff substrate. A more complete treatment, allowing substrate contraction, would be much more computationally intensive. Finally, a last set of results concerns simulating a recent experiment where cells were constrained to move along an annulus. As an initially small cluster expands, the two ends eventually collide and the cluster transitions to the coherent motion state with almost all the cells having the same polarity and no division taking place. We observe that this transition takes place by a polarity reversing wave that eventually leads to a large preponderance of cells moving the same way. Again, wave-like phenomena have been seen in colliding tissues [42]. There can be individual "rebellious cells" that maintain the "wrong" polarity, but these have little effect on the overall cluster behavior. Measurements of individual cell polarizations, necessary for showing the existence of rebellious cells, might be accomplished by determining the relative position of intracellular structures such as centrosomes [43].

There are a number of directions in which our model could be extended. It should be straightforward in principle to create a two-dimensional version of our system. This extension has already been accomplished for a contraction-protrusion model of single cell motility, with the major changes being that now both force and torque need to be balanced at each step of the simulation and the fact that polarity now becomes a vector which determines the direction of the protrusion [35]. A different generalization involves incorporating the effects of substrate elasticity, as discussed above. Given that the entire measurement strategy involved in traction force microscopy relies on having a flexible substrate, it is clearly important to understand when this feature makes a real difference. In addition, As we have seen, there is clear evidence in favor of collective durotaxis, namely that the cluster can respond to relatively small substrate stiffness gradients, sufficiently small as to preclude single-cell durotaxis. We have shown phenomenologically how making the spring constant or the detachment force depend on position can lead to durotaxis, but it could be useful to replace this approach with one that explicitly accounts for substrate elasticity, as well as including possible parameter changes at the single-cell level due to some form of mechanosensing.

Cells are extremely complex mechanical objects and of course one cannot expect to describe all their phenomenology with simple models. However, at least for collective behavior we may expect (or at least hope) that many of the biological details are not critical when it comes to grasping the essence of what can occur. The results reported here should give us added confidence in this physics-based approach.

# Acknowledgements

This work was supported in part by the National Science Foundation Center for Theoretical Biological Physics NSF PHY-2019745, and also by PHY-1605817 and NSF-EFRI-1741618.

# References

[1] Thomas D Pollard and Gary G Borisy. Cellular motility driven by assembly and disassembly of actin filaments. Cell, 112(4):453–465, 2003.

[2] Peter Friedl, Joseph Locker, Erik Sahai, and Jeffrey E Segall. Classifying collective cancer cell invasion. Nature cell biology, 14(8):777–783, 2012.

[3] Vincent Hakim and Pascal Silberzan. Collective cell migration: a physics perspective. Reports On Progress In Physics, 80(7):076601–47, April 2017.

[4] Celina G Kleer, Kenneth L van Golen, Thomas Braun, and Sofia D Merajver. Persistent E-Cadherin Expression in Inflammatory Breast Cancer. Modern Pathology, 14(5):458–464, May 2001.

[5] Mohit Kumar Jolly, Marcelo Boareto, Bisrat G Debeb, Nicola Aceto, Mary C Farach-Carson, Wendy A Woodward, and Herbert Levine. Inflammatory breast cancer: a model for investigating cluster-based dissemination. NPJ Breast Cancer, 3(1):1–8, 2017.

[6] M Cristina Marchetti, Jean-François Joanny, Sriram Ramaswamy, Tanniemola B Liverpool, Jacques Prost, Madan Rao, and R Aditi Simha. Hydrodynamics of soft active matter. Reviews of Modern Physics, 85(3):1143, 2013.

[7] Alpha S Yap, William M Brieher, and Barry M Gumbiner. Molecular and functional analysis of cadherin-based adherens junctions. Annual review of cell and developmental biology, 13(1):119–146, 1997.

[8] Roberto Mayor and Carlos Carmona-Fontaine. Keeping in touch with contact inhibition of locomotion. Trends in cell biology, 20(6):319–328, 2010.

[9] Xavier Trepat and Erik Sahai. Mesoscale physical principles of collective cell organization. Nature Physics, 14(7):671–682, 2018.

[10] Xavier Trepat, Michael R Wasserman, Thomas E Angelini, Emil Millet, David A Weitz, James P Butler, and Jeffrey J Fredberg. Physical forces during collective cell migration. Nature physics, 5(6):426–430, 2009.

[11] Raimon Sunyer, Vito Conte, Jorge Escribano, Alberto Elosegui-Artola, Anna Labernadie, Léo Valon, Daniel Navajas, José Manuel García-Aznar, José J Muñoz, Pere Roca-Cusachs, and Xavier Trepat. Collective cell durotaxis emerges from long-range intercellular force transmission. Science (New York, NY), 353(6304):1157–1161, September 2016.

[12] Xavier Serra-Picamal, Vito Conte, Romaric Vincent, Ester Anon, Dhananjay T Tambe, Elsa Bazellieres, James P Butler, Jeffrey J Fredberg, and Xavier Trepat. Mechanical waves during tissue expansion. Nature Physics, 8(8):628–634, July 2012.

[13] Sham Tlili, Estelle Gauquelin, Brigitte Li, Olivier Cardoso, Benoît Ladoux, Hélène Delanoë-Ayari, and François Graner. Collective cell migration without proliferation: density determines cell velocity and wave velocity. Royal Society Open Science, 5(5):172421–20, May 2018.

[14] Shreyansh Jain, Victoire M L Cachoux, Gautham H N S Narayana, Simon de Beco, Joseph D'Alessandro, Victor Cellerin, Tianchi Chen, Mélina L Heuzé, Philippe Marcq, René-Marc Mège, Alexandre J Kabla, Chwee Teck Lim, and Benoît Ladoux. The role of single-cell mechanical behaviour and polarity in driving collective cell migration. Nature Physics, 10:1–8, May 2020.

[15] C M Lo, H B Wang, M Dembo, and Y L Wang. Cell movement is guided by the rigidity of the substrate. Biophysical Journal, 79(1):144–152, June 2000.

[16] Elizaveta A Novikova, Matthew Raab, Dennis E Discher, and Cornelis Storm. Persistence-driven durotaxis: Generic, directed motility in rigidity gradients. Physical review letters, 118(7):078103, 2017.

[17] Guangyuan Yu, Jingchen Feng, Haoran Man, and Herbert Levine. Phenomenological modeling of durotaxis. Physical Review E, 96(1):010402, 2017.

[18] Charles R Doering, Xiaoming Mao, and Leonard M Sander. Random walker models for durotaxis. Physical biology, 15(6):066009–8, November 2018.

[19] Jorge Escribano, Raimon Sunyer, María Teresa Sánchez, Xavier Trepat, Pere Roca-Cusachs, and José Manuel García-Aznar. A hybrid computational model for collective cell durotaxis. Biomechanics And Modeling In Mechanobiology, 17(4):1037–1052, March 2018.

[20] C Blanch-Mercader, R Vincent, E Bazellières, X Serra-Picamal, X Trepat, and J Casademunt. Effective viscosity and dynamics of spreading epithelia: a solvable model. Soft Matter, 13(6):1235–1243, 2017.

[21] Shiladitya Banerjee, Kazage J C Utuje, and M Cristina Marchetti. Propagating Stress Waves During Epithelial Expansion. Physical Review Letters, 114(22):228101–5, June 2015.

[22] Juliane Zimmermann, Brian A Camley, Wouter-Jan Rappel, and Herbert Levine. Contact inhibition of locomotion determines cell–cell and cell–substrate forces in tissues. Proceedings of the National Academy of Sciences of the United States of America, 113(10):2660–2665, March 2016.

[23] John J Williamson and Guillaume Salbreux. Stability and roughness of interfaces in mechanically regulated tissues. Physical review letters, 121(23):238102, 2018.

[24] Ricard Alert, Carles Blanch-Mercader, and Jaume Casademunt. Active fingering instability in tissue spreading. Physical review letters, 122(8):088104, 2019.

[25] Yanjun Yang and Herbert Levine. Leader-cell-driven epithelial sheet fingering. Physical Biology, 0(0):0, May 2020.

[26] François Graner and James A Glazier. Simulation of biological cell sorting using a two-dimensional extended potts model. Physical review letters, 69(13):2013, 1992.

[27] Reza Farhadifar, Jens-Christian Röper, Benoit Aigouy, Suzanne Eaton, and Frank Jülicher. The influence of cell mechanics, cell-cell interactions, and proliferation on epithelial packing. Current Biology, 17(24):2095–2104, 2007.

[28] Brian A Camley, Yunsong Zhang, Yanxiang Zhao, Bo Li, Eshel Ben-Jacob, Herbert Levine, and Wouter-Jan Rappel. Polarity mechanisms such as contact inhibition of locomotion regulate persistent rotational motion of mammalian cells on micropatterns. Proceedings of the National Academy of Sciences, 111(41):14770–14775, 2014.

[29] Sara Najem and Martin Grant. Phase-field model for collective cell migration. Physical Review E, 93(5):052405, 2016.

[30] Tamás Vicsek, András Czirók, Eshel Ben-Jacob, Inon Cohen, and Ofer Shochet. Novel type of phase transition in a system of self-driven particles. Physical review letters, 75(6):1226, 1995.

[31] Markus Basan, Jens Elgeti, Edouard Hannezo, Wouter-Jan Rappel, and Herbert Levine. Alignment of cellular motility forces with tissue flow as a mechanism for efficient wound healing. Proceedings of the National Academy of Sciences, 110(7):2452–2459, 2013.

[32] Benjamin L Bangasser, Ghaidan A Shamsan, Clarence E Chan, Kwaku N Opoku, Erkan T uuml zel, Benjamin W Schlichtmann, Jesse A Kasim, Benjamin J Fuller, Brannon R McCullough, Steven S Rosenfeld, and David J Odde. Shifting the optimal stiffness for cell migration. Nature Communications, 8:1–10, May 2017.

[33] Norikazu Yamana, Yoshiki Arakawa, Tomohiro Nishino, Kazuo Kurokawa, Masahiro Tanji, Reina E Itoh, James Monypenny, Toshimasa Ishizaki, Haruhiko Bito, Kazuhiko Nozaki, et al. The rho-mdia1 pathway regulates cell polarity and focal adhesion turnover in migrating cells through mobilizing apc and c-src. Molecular and cellular biology, 26(18):6844–6858, 2006.

[34] Mathias Buenemann, Herbert Levine, Wouter-Jan Rappel, and Leonard M Sander. The Role of Cell Contraction and Adhesion in Dictyostelium Motility. Biophysical Journal, 99(1):50–58, July 2010.

[35] Jingchen Feng, Herbert Levine, Xiaoming Mao, and Leonard M Sander. Cell motility, contact guidance, and durotaxis. Soft Matter, 15:4856–4864, June 2019.

[36] Yanjun Yang and Herbert Levine. Role of the supracellular actomyosin cable during epithelial wound healing. Soft Matter, 14(23):4866–4873, 2018.

[37] Sebastian J Streichan, Christian R Hoerner, Tatjana Schneidt, Daniela Holzer, and Lars Hufnagel. Spatial constraints control cell proliferation in tissues. Proceedings of the National Academy of Sciences, 111(15):5586–5591, 2014.

[38] Robert J Pelham and Yu-li Wang. Cell locomotion and focal adhesions are regulated by substrate flexibility. Proceedings of the National Academy of Sciences, 94(25):13661–13665, 1997.

[39] Ricard Alert and Jaume Casademunt. Role of Substrate Stiffness in Tissue Spreading: Wetting Transition and Tissue Durotaxis. Langmuir, 35(23):7571–7577, October 2018.

[40] Falko Ziebert and Igor S Aranson. Effects of adhesion dynamics and substrate compliance on the shape and motility of crawling cells. PloS one, 8(5):e64511, 2013.

[41] Tamal Das, Kai Safferling, Sebastian Rausch, Niels Grabe, Heike Boehm, and Joachim P Spatz. A molecular mechanotransduction pathway regulates collective migration of epithelial cells. Nature cell biology, 17(3):276–287, 2015.

[42] Pilar Rodríguez-Franco, Agustí Brugués, Ariadna Marín-Llauradó, Vito Conte, Guiomar Solanas, Eduard Batlle, Jeffrey J Fredberg, Pere Roca-Cusachs, Raimon Sunyer, and Xavier Trepat. Long-lived force patterns and deformation waves at repulsive epithelial boundaries. Nature materials, 16(10):1029–1037, 2017.

[43] N Tang and W F Marshall. Centrosome positioning in vertebrate development. Journal of cell science, 125(21):4951–4961, December 2012.

# Collective Motility, Mechanical Waves, and Durotaxis in Cell Clusters
# Supplemental Information

Youyuan Deng[1,2], Herbert Levine[1,3], Xiaoming Mao[4], and Leonard M. Sander[4,5]

[1]Center for Theoretical Biological Physics, Rice University, Houston, Texas 77030-1402, USA
[2]Applied Physics Graduate Program, Rice University, Houston, Texas 77005-1827, USA
[3]Department of Physics, Northeastern University, Boston, Massachusetts 02115, USA
[4]Department of Physics, University of Michigan, Ann Arbor, Michigan 48109-1040, USA
[5]Center for the Study of Complex Systems, University of Michigan, Ann Arbor, Michigan 48109-1107, USA

January 16, 2021

## 1 Contact Inhibition Of Locomotion, Two-Cell Demonstration

Figs 1, 2 respectively are two cases of one and both cells flipping after a head-to-head collision.

## 2 Random Initial Polarity and CIL

Running the simulations without division but with random initial polarity gives a variety of different final states. We hereby show that a larger cluster is more likely to end up in configurations with two domains in a static tug-of-war and no bulk translation. As we pointed out, a model cell cluster will eventually end up comprising of two domains, the left domain of cells polarizing left, and a corresponding right domain. This process of CIL can be abstracted into the following scheme: for a fixed number of cells (a fixed colony size), given random initial polarity of each cell, for each step, randomly select a cell (say $i$-th cell), and compare its polarity with the neighbor next to its head ($i'$-th cell, $i' = i \pm 1$); randomly flip one of the pair if they are in head-to-head collision. Then, for many such samples, count the relative portion of the left domain, which is the same as the "domain wall" relative position. As seen from Fig 3, the probability distribution is nearly unchanged with varied colony size. Note that we need a minimum absolute number of unstalled edge cells in each domain to balance the traction generated at the opposite edge and to stall the interior cells, or there would be bulk translation. Namely, each domain must have a minimum absolute number of cells. For a larger colony, the same minimum absolute number corresponds to a lower minimum left/right domain portion. Thus, it corresponds to a higher probability of being static for larger colonies.

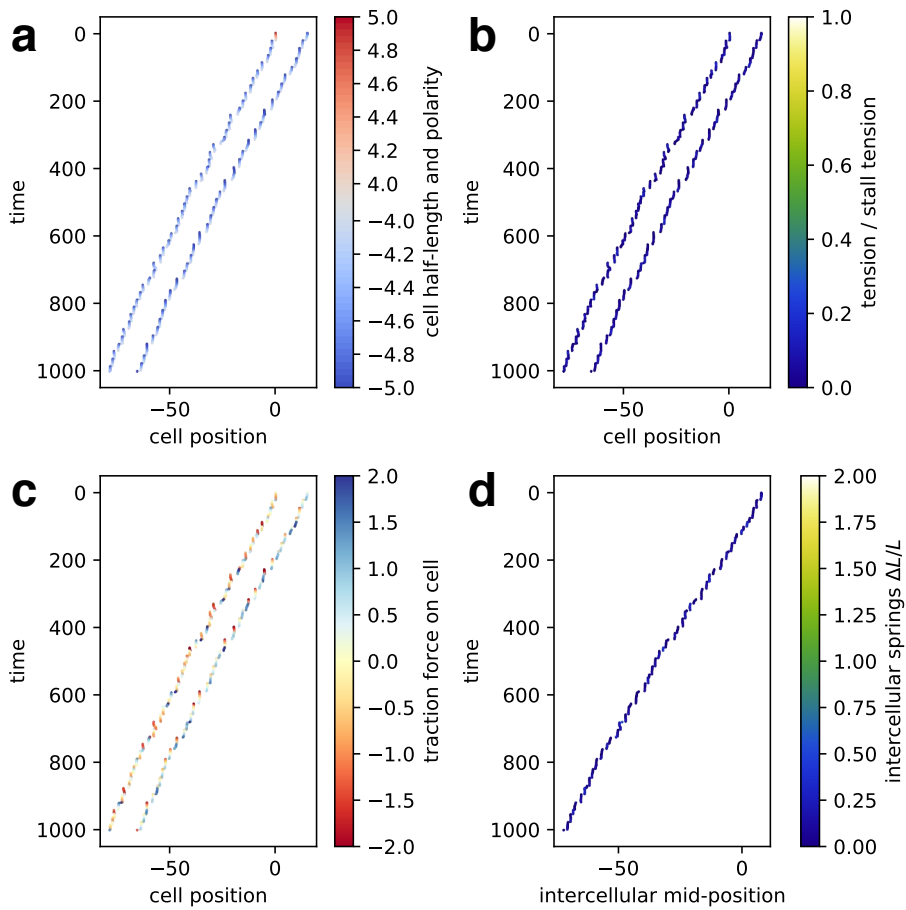

Figure 1: Space-time plot of two cells initially aligned head-to-head. One cell flips polarity. (a) Polarity and half-length of the cells. Blue or negative sign denotes "left" polarity and red/positive denotes "right". (b) Cell midpoint tension. (c) Traction force on each cell by substrate due to the adhesions. (d) Inter-cellular tension stress, i.e. the stretch of the inter-cellular springs.

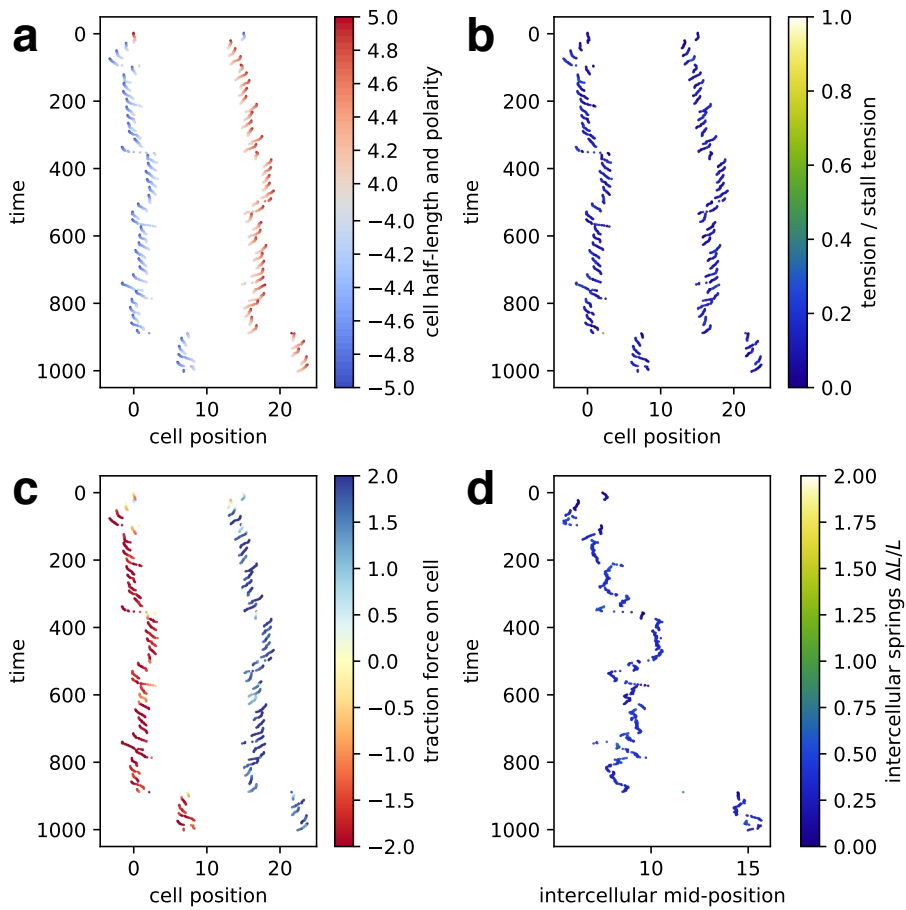

Figure 2: Space-time plot of two cells initially aligned head-to-head. Both cells flip polarity. (a) Polarity and half-length of the cells. Blue or negative sign denotes "left" polarity and red/positive denotes "right". (b) Cell midpoint tension. (c) Traction force on each cell by substrate due to the adhesions. (d) Inter-cellular tension stress, i.e. the stretch of the inter-cellular springs.

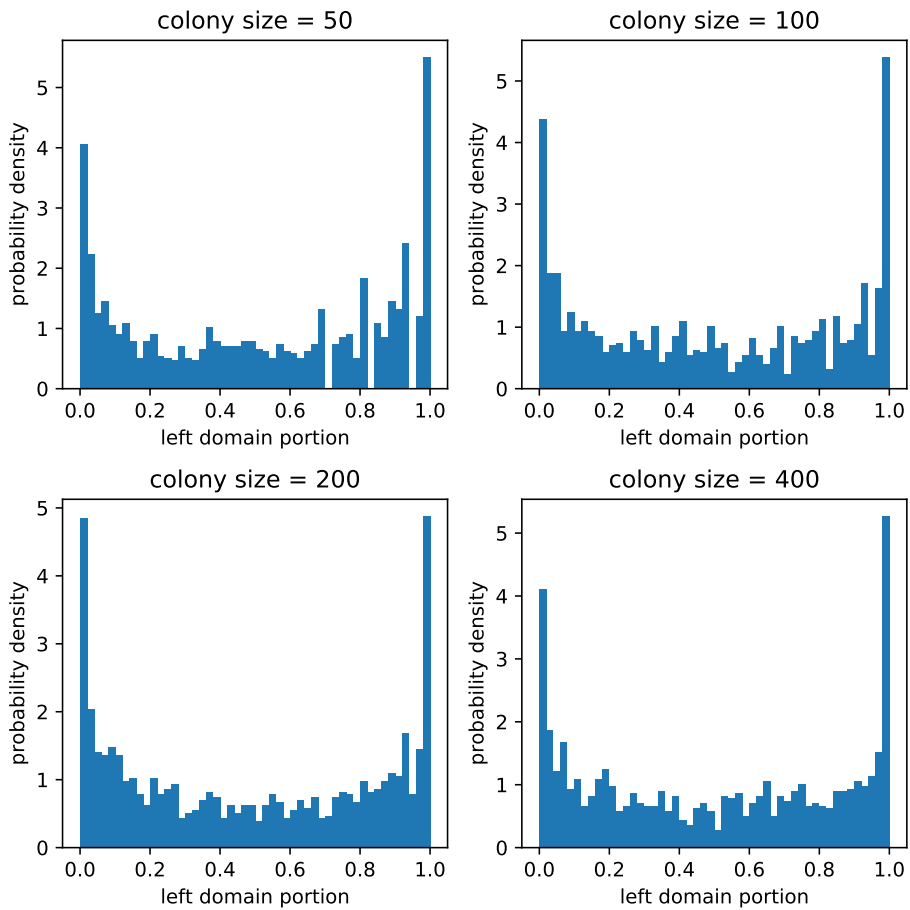

Figure 3: Final polarity outcome: probability distribution over the portion of left domain (number of cells in left domain divided by total cell number). Frequency counted from 1280 samples for each of the colony sizes.

## 3 Cell Division Rate and Inter-cellular Tension

In main text Fig 4, the influence of division rate on intra-cellular tension is shown. The influence on inter-cellular tension is qualitatively similar; see Fig 4.

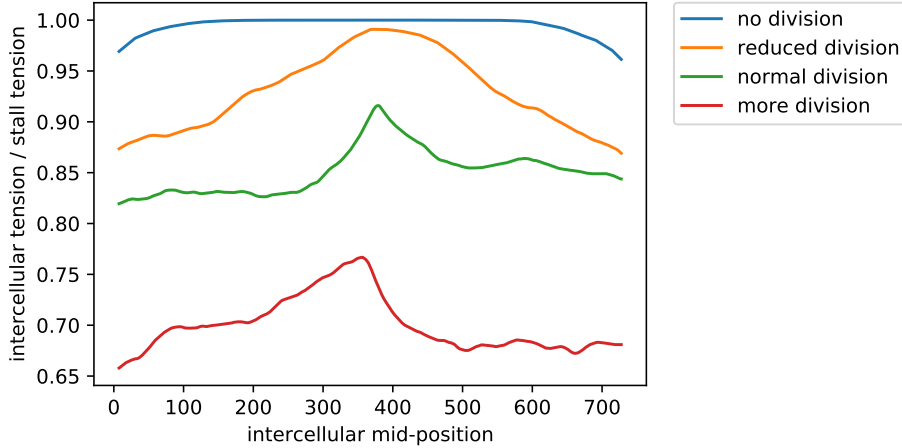

Figure 4: Inter-cellular tension averaged over the latter half of a trajectory for different division frequency. Reduced division: $k_{div} = 0.005$, $T_{div} = 0.99T_s$; Normal division (same as in the parameter table): $k_{div} = 1$, $T_{div} = 0.99T_s$; More division: $k_{div} = 1$, $T_{div} = 0.9T_s$.

## 4 Larger Spring Constant Of Cell-Substrate Adhesions Leads To Faster Cell Speed

As mentioned in the main text, larger $k$ should lead to faster cell speed since it makes detachment faster. In Fig 5, for each value of $k$, we simulated 50 samples of free single cell and calculated the mean / standard deviation of each sample's average speed. A positive correlation is clear. Larger $k$ as an approximation of larger local stiffness leads to faster speed. It is therefore consistent with the proposed mechanism that faster cell speed on stiffer substrates leads to population durotaxis [1, 2].

## 5 Wave Speed Numerical Experiments

In main text Fig 4, we did several measurements on the two types of waves, using varied model parameters. For the release wave (Fig 4(f)), changed parameters come from the Cartesian product of $\{v_f = 5, 3, 7\}$, $\{F_{hs} = 1.5, 1.0, 2.0\}$, $\{l_0 = 10, 8, 12\}$. For the division wave (Fig 4(g)), changed parameters come from the Cartesian product of $\{v_f = 5, 2.5, 10\}$, $\{\Delta x_c^{(init)} = 15, 20, 30\}$. $\Delta x_c^{(init)}$ is the initial separation between adjacent cell midpoints; $l_0 = \Delta x_c^{(init)} - 2L_0$; $L_0$ is fixed. i.e.

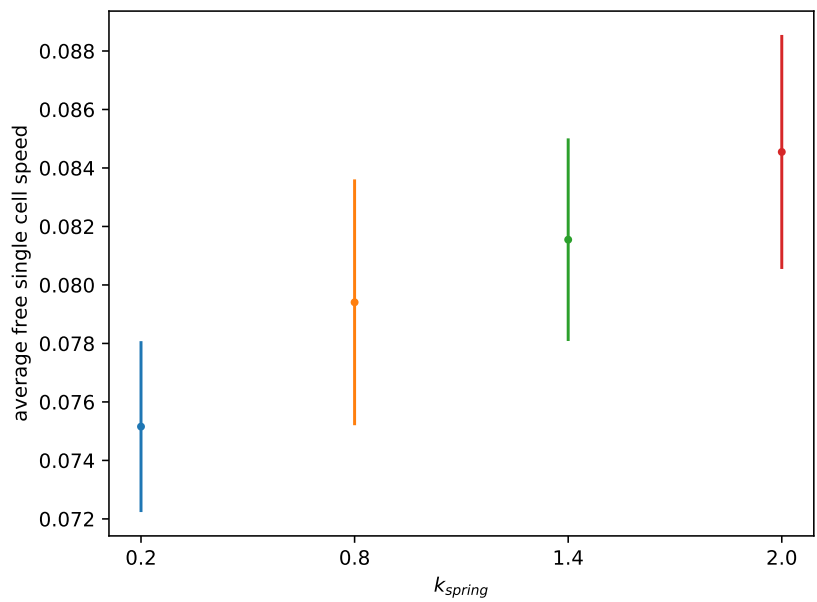

Figure 5: Average cell speed vs $k$, the spring constant of cell-substrate adhesions. Error bar is standard deviation.

Compression due to confinement is not included. See the parameters table for symbol notation. The other parameters are kept the default values as in the table.

# 6 More Kymographs

There are some kymographs supplementing examples in the main text. See their respective captions for details.

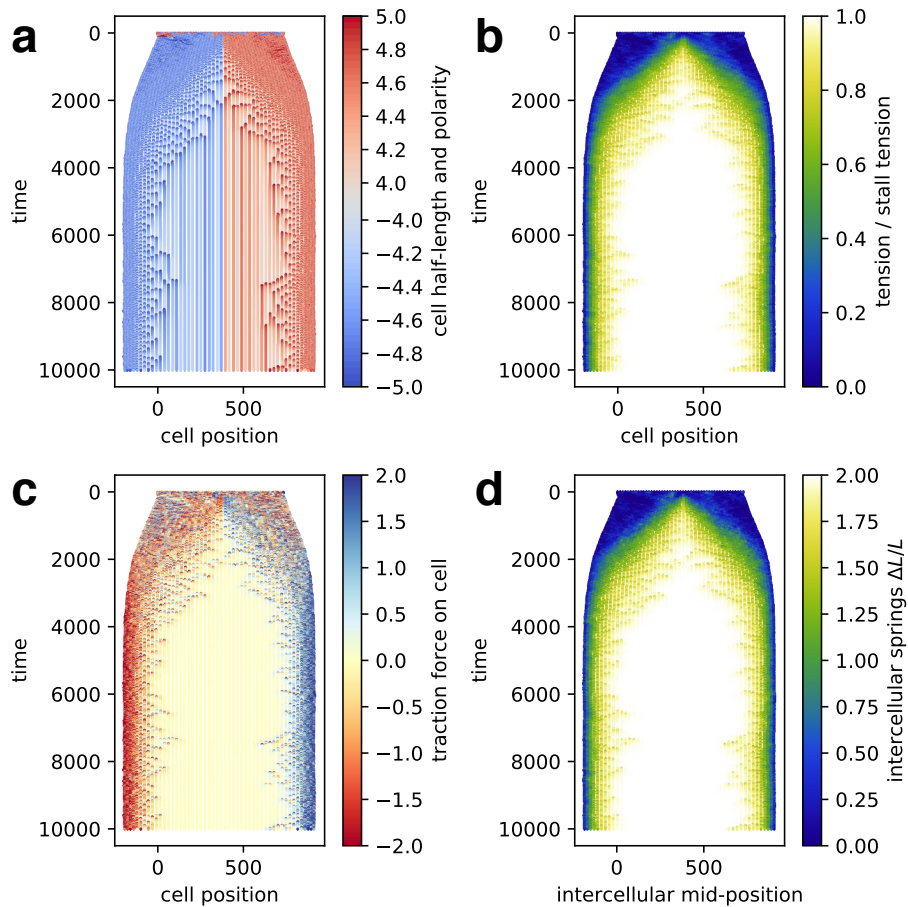

Figure 6: Space-time plot of the sample of main text Fig 2. (a) Polarity and half-length of the cells. Blue or negative sign denotes "left" polarity and red/positive denotes "right". (b) Cell midpoint tension. (c) Traction force on each cell by substrate due to the adhesions. (d) Inter-cellular tension stress, i.e. the stretch of the inter-cellular springs.

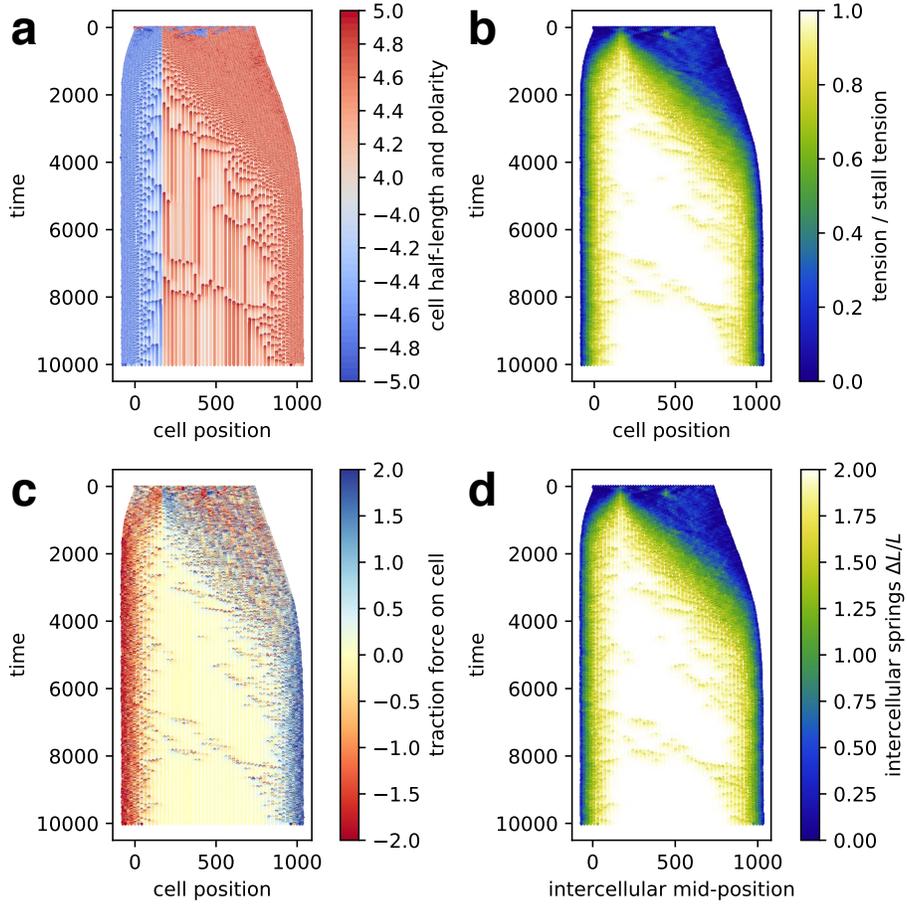

Figure 7: Space-time plot of the sample of main text Fig 3(a-c). (a) Polarity and half-length of the cells. Blue or negative sign denotes "left" polarity and red/positive denotes "right". (b) Cell midpoint tension. (c) Traction force on each cell by substrate due to the adhesions. (d) Inter-cellular tension stress, i.e. the stretch of the inter-cellular springs.

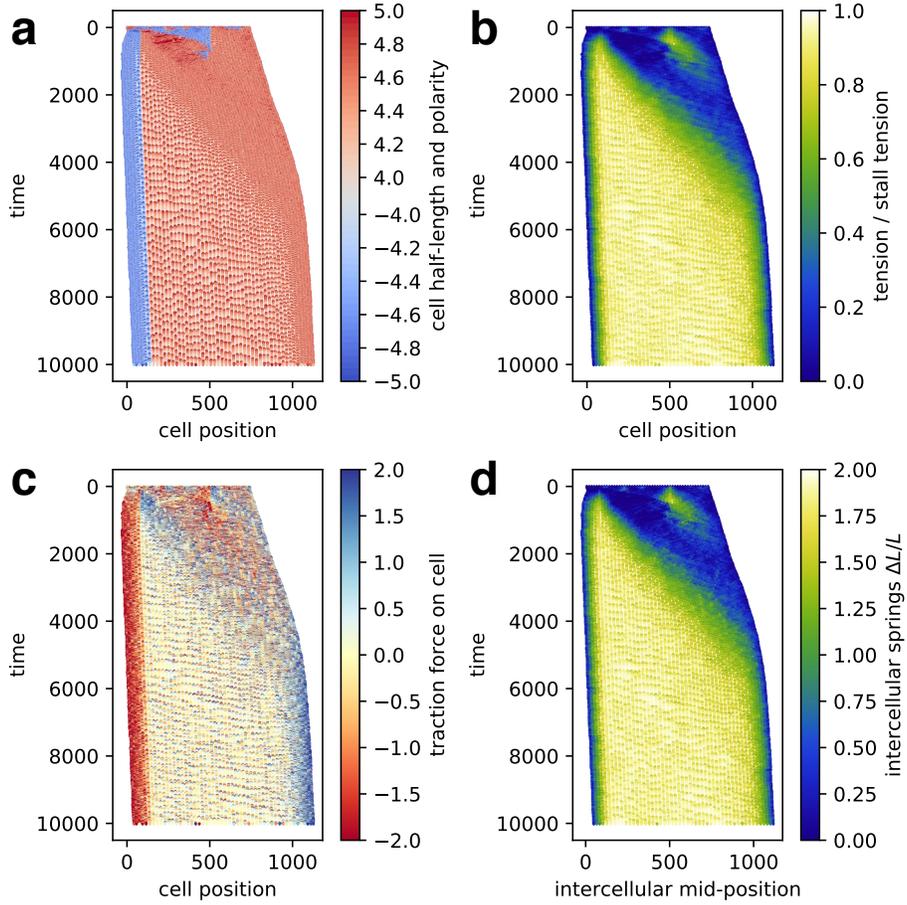

Figure 8: Space-time plot of the sample of main text Fig 3(d-f). (a) Polarity and half-length of the cells. Blue or negative sign denotes "left" polarity and red/positive denotes "right". (b) Cell midpoint tension. (c) Traction force on each cell by substrate due to the adhesions. (d) Inter-cellular tension stress, i.e. the stretch of the inter-cellular springs.

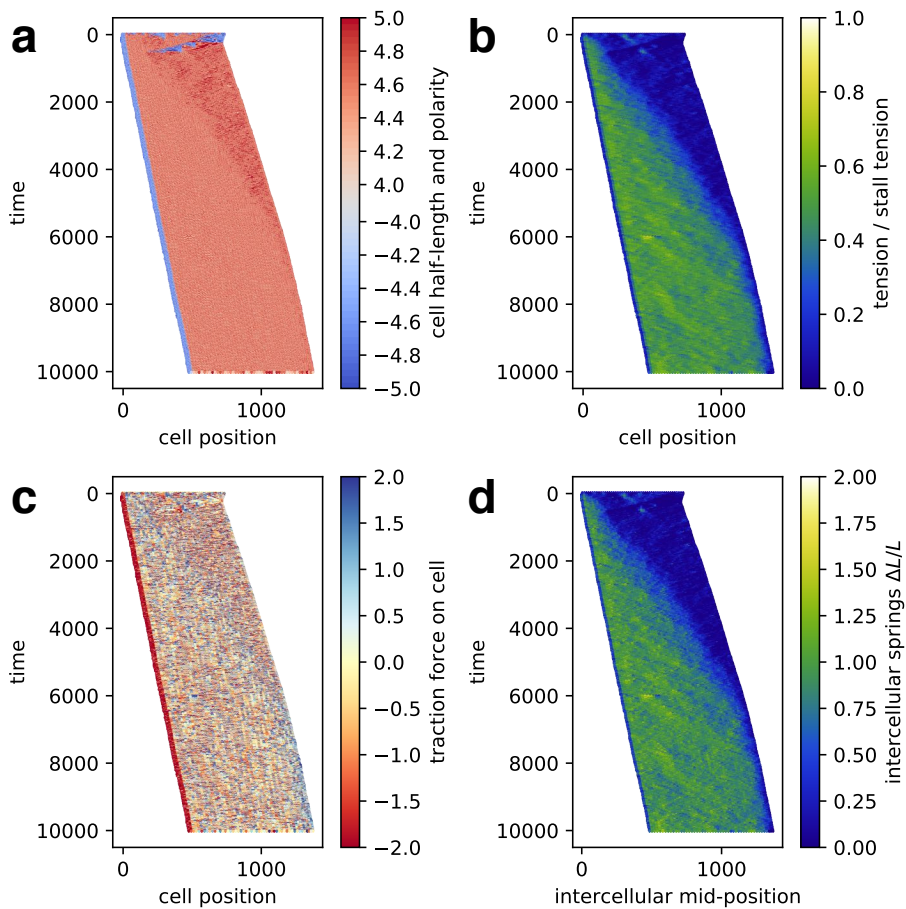

Figure 9: Space-time plot of a sample without proliferation that ends up having a very thin domain on one side. (a) Polarity and half-length of the cells. Blue or negative sign denotes "left" polarity and red/positive denotes "right". (b) Cell midpoint tension. (c) Traction force on each cell by substrate due to the adhesions. (d) Inter-cellular tension stress, i.e. the stretch of the inter-cellular springs.

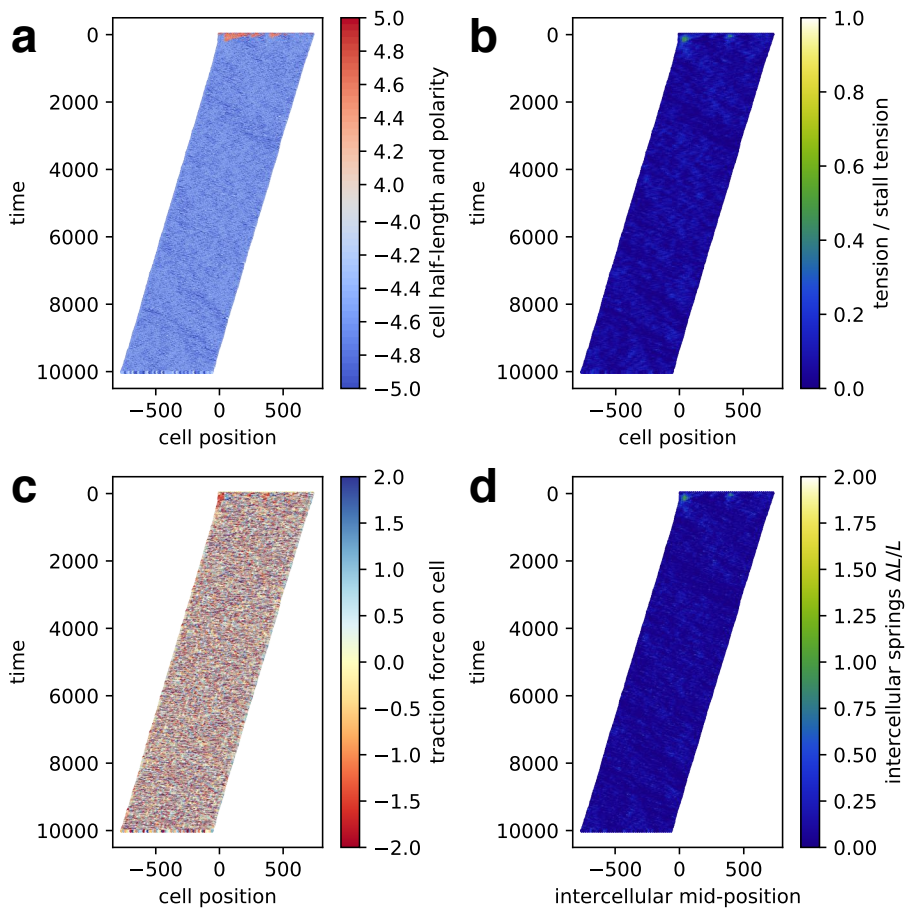

Figure 10: Space-time plot of a sample without proliferation that ends up having all cells of the same polarity. (a) Polarity and half-length of the cells. Blue or negative sign denotes "left" polarity and red/positive denotes "right". (b) Cell midpoint tension. (c) Traction force on each cell by substrate due to the adhesions. (d) Inter-cellular tension stress, i.e. the stretch of the inter-cellular springs.

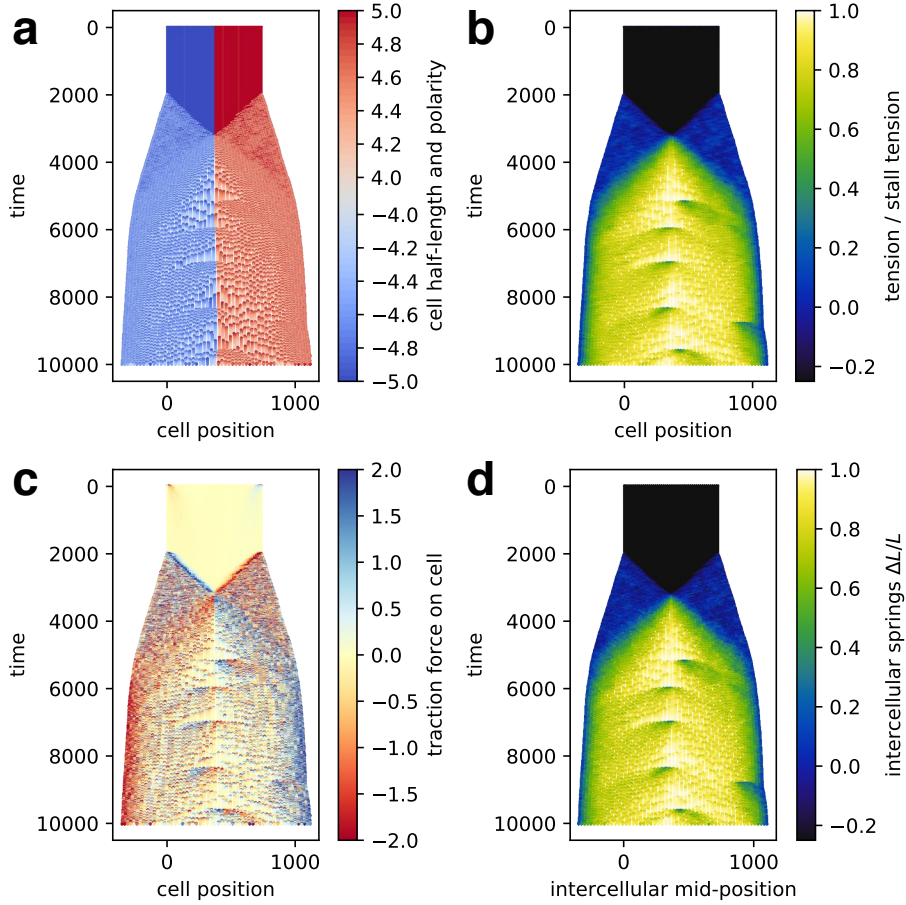

Figure 11: Space-time plot of the sample of main text Fig 4(bc). (a) Polarity and half-length of the cells. Blue or negative sign denotes "left" polarity and red/positive denotes "right". (b) Cell midpoint tension. (c) Traction force on each cell by substrate due to the adhesions. (d) Inter-cellular tension stress, i.e. the stretch of the inter-cellular springs.

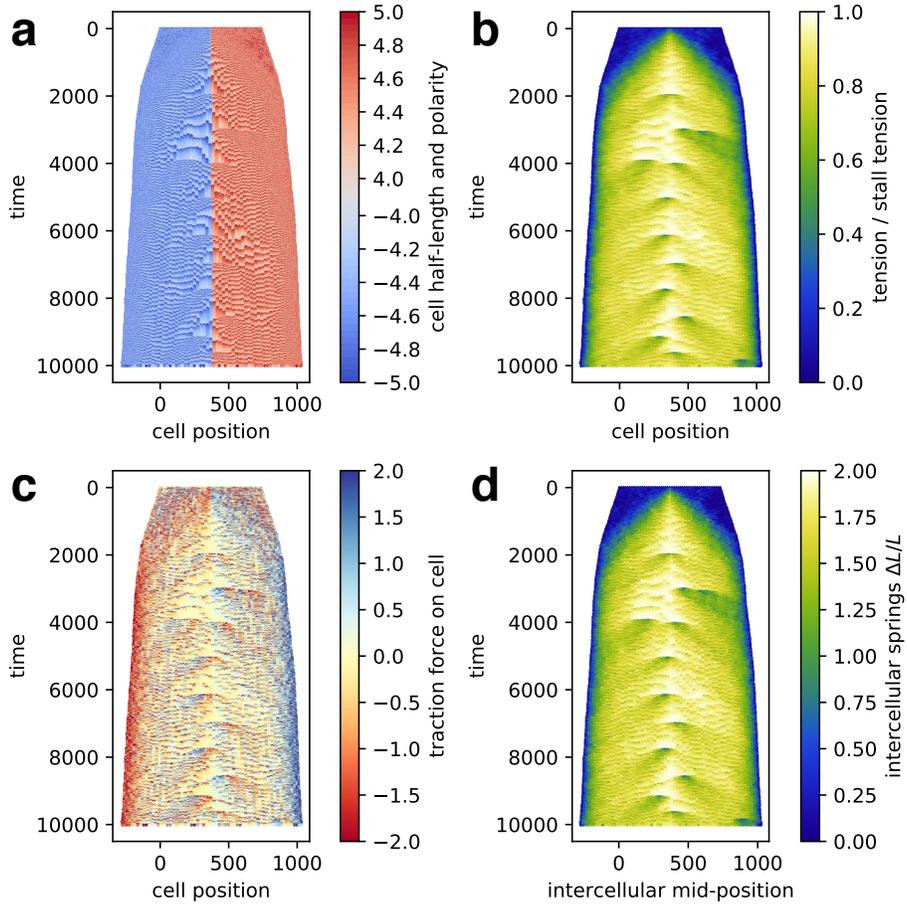

Figure 12: Space-time plot of the sample with normal division rate (the same parameters as in the parameters table) mentioned in main text Fig 4. (a) Polarity and half-length of the cells. Blue or negative sign denotes "left" polarity and red/positive denotes "right". (b) Cell midpoint tension. (c) Traction force on each cell by substrate due to the adhesions. (d) Inter-cellular tension stress, i.e. the stretch of the inter-cellular springs.

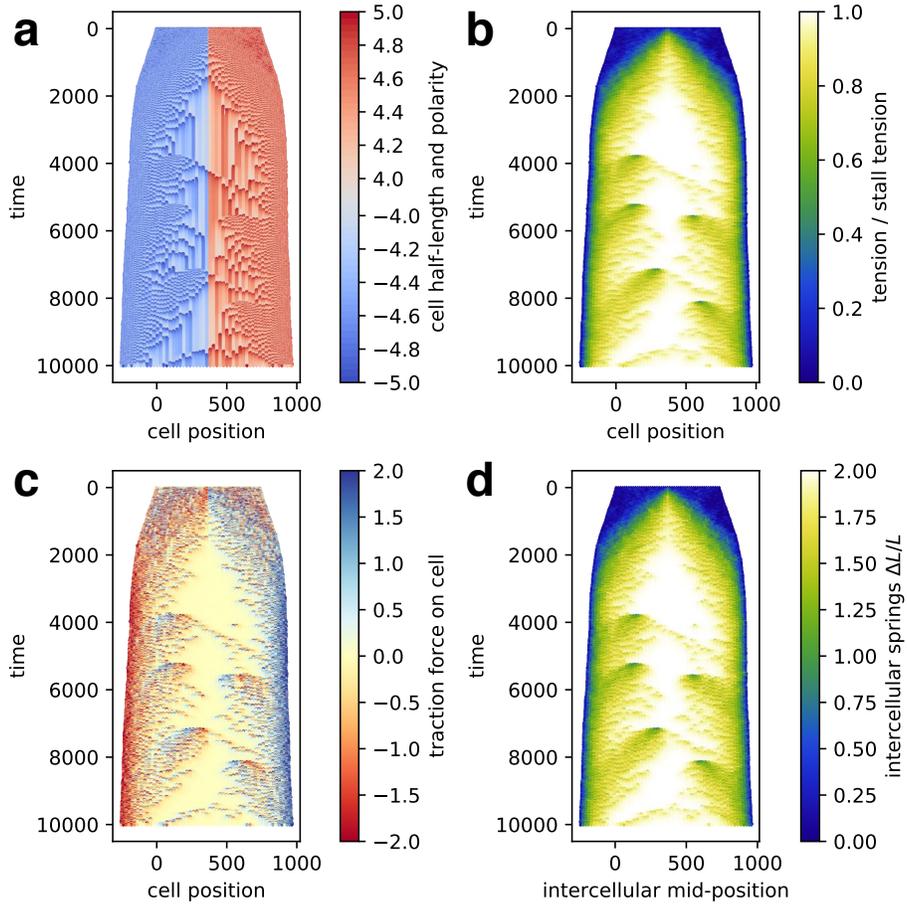

Figure 13: Space-time plot of the sample of main text Fig 4(d), i.e. the sample with reduced division rate. (a) Polarity and half-length of the cells. Blue or negative sign denotes "left" polarity and red/positive denotes "right". (b) Cell midpoint tension. (c) Traction force on each cell by substrate due to the adhesions. (d) Inter-cellular tension stress, i.e. the stretch of the inter-cellular springs.

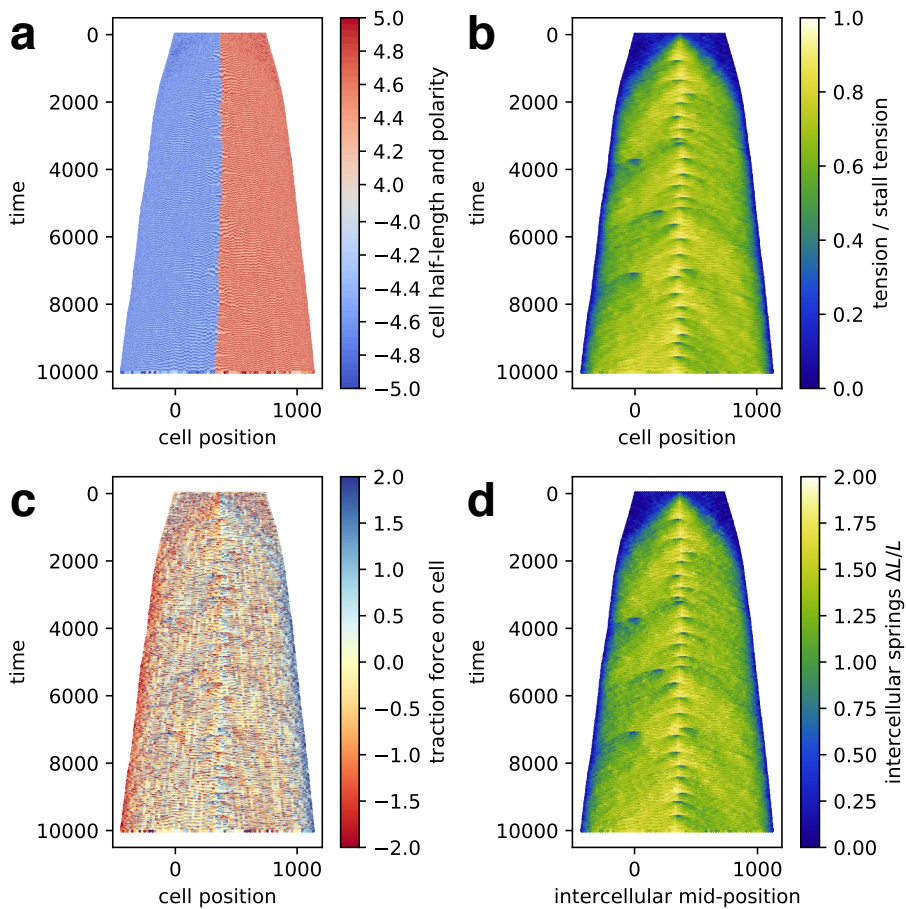

Figure 14: Space-time plot of the sample of main text Fig 4(e), i.e. the sample with increased division rate. (a) Polarity and half-length of the cells. Blue or negative sign denotes "left" polarity and red/positive denotes "right". (b) Cell midpoint tension. (c) Traction force on each cell by substrate due to the adhesions. (d) Inter-cellular tension stress, i.e. the stretch of the inter-cellular springs.

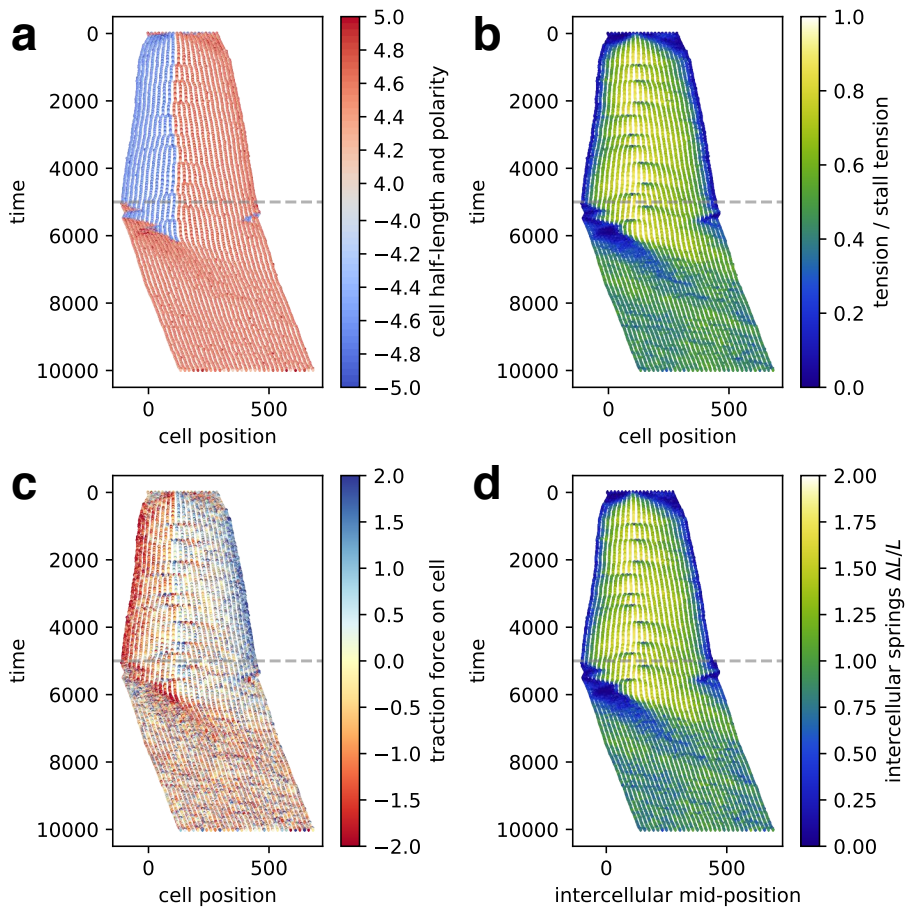

Figure 15: Space-time plot of a cell cluster that keeps expanding until two outermost cells are joined by a spring at $t = 5000$, indicated by the gray dashed line. A sample with a somewhat smooth polarity border. (a) Polarity and half-length of the cells. Blue or negative sign denotes "left" polarity and red/positive denotes "right". (b) Cell midpoint tension. (c) Traction force on each cell by substrate due to the adhesions. (d) Inter-cellular tension stress, i.e. the stretch of the inter-cellular springs. Note that the rightmost new spring created at $t = 5000$ is the one connecting two outermost cells.

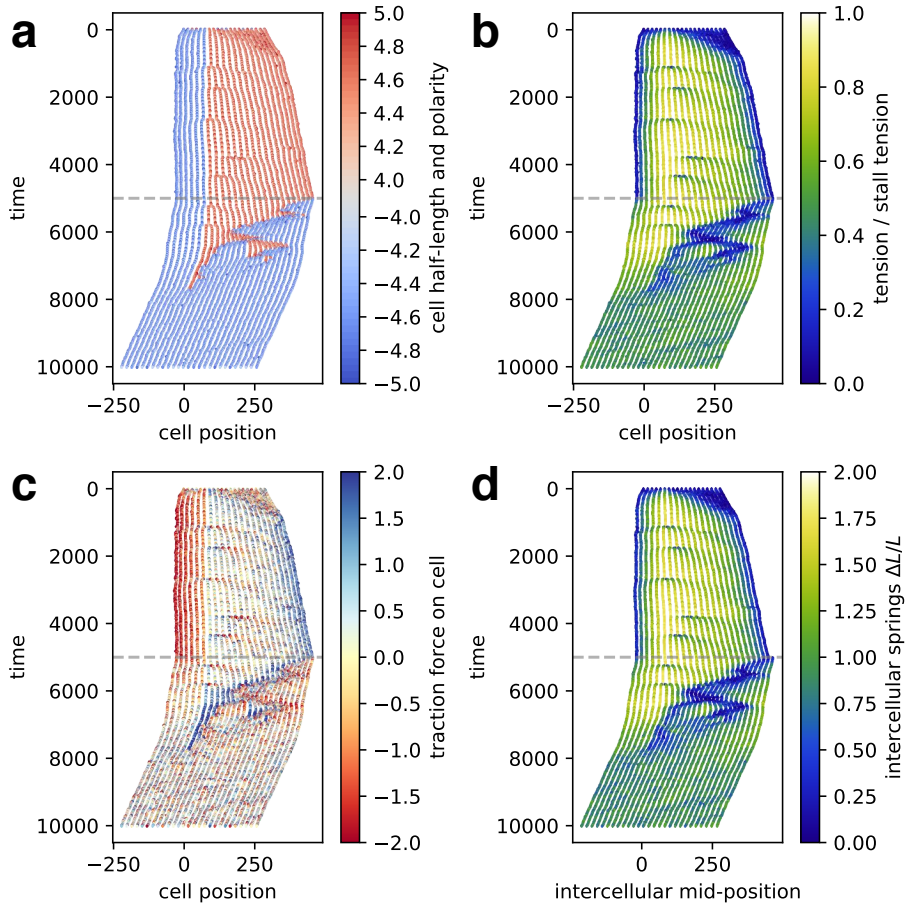

Figure 16: Space-time plot of a cell cluster that keeps expanding until two outermost cells are joined by a spring at $t = 5000$, indicated by the gray dashed line. A sample with a zig-zag polarity border. (a) Polarity and half-length of the cells. Blue or negative sign denotes "left" polarity and red/positive denotes "right". (b) Cell midpoint tension. (c) Traction force on each cell by substrate due to the adhesions. (d) Inter-cellular tension stress, i.e. the stretch of the inter-cellular springs. Note that the rightmost new spring created at $t = 5000$ is the one connecting two outermost cells.

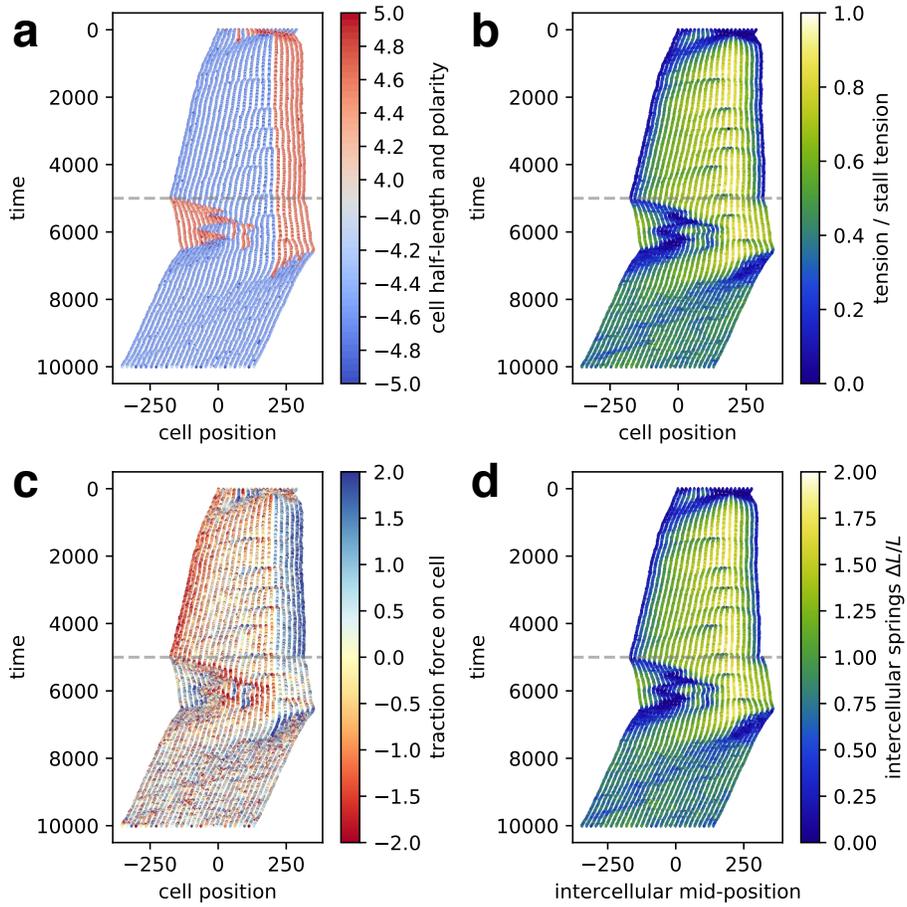

Figure 17: Space-time plot of a cell cluster that keeps expanding until two outermost cells are joined by a spring at $t = 5000$, indicated by the gray dashed line. A sample with a zig-zag polarity border. (a) Polarity and half-length of the cells. Blue or negative sign denotes "left" polarity and red/positive denotes "right". (b) Cell midpoint tension. (c) Traction force on each cell by substrate due to the adhesions. (d) Inter-cellular tension stress, i.e. the stretch of the inter-cellular springs. Note that the rightmost new spring created at $t = 5000$ is the one connecting two outermost cells.

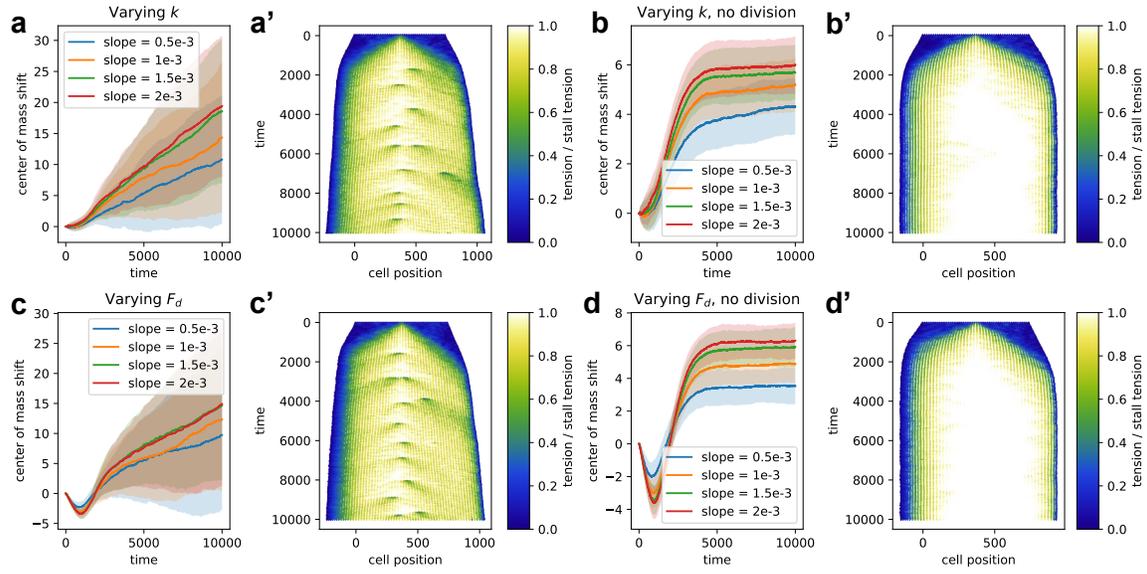

Figure 18: Durotaxis: The necessity of cell division. (a–d) Center of mass shift of the clusters as a function of time, for two effective stiffness mechanisms and different gradients of stiffness. Each line is the average over 100 samples, and the filled area signifies the standard deviation. These are the same plots as main text Fig 6. (a)(b) Space-dependent elastic constant $k$: $k = \max(0.2, 0.4 + \text{slope} * x_c)$, with proliferation on (a) / off (b). (c)(d) Space-dependent $F_d$: $F_d = \max(0.2, 0.4 + \text{slope} * x_c)$, with proliferation on (c) / off (d). (a'–d') Cell midpoint tension kymographs of a sample with slope $1.5e-3$ from each of the groups (a–d), the same samples as in main text Fig 6.

## Movies

**S1.** Animation of a free moving cell.

**S2.** Animation of a pair of colliding cell where one cell flips polarity.

**S3.** Animation of a pair of colliding cell where both cells flip polarity.

## Methods

The simulation is performed by a custom-built program. We utilized the optimization function (for mechanical equilibration) and others from DLIB [3].

## References

[1] Elizaveta A Novikova, Matthew Raab, Dennis E Discher, and Cornelis Storm. Persistence-driven durotaxis: Generic, directed motility in rigidity gradients. Physical review letters, 118(7):078103, 2017.

[2] Guangyuan Yu, Jingchen Feng, Haoran Man, and Herbert Levine. Phenomenological modeling of durotaxis. Physical Review E, 96(1):010402, 2017.

[3] Davis E. King. Dlib-ml: A machine learning toolkit. Journal of Machine Learning Research, 10:1755–1758, 2009.